 \documentclass[twocolumn]{aastex6}
 \usepackage{amsmath}

\usepackage{color}

\fullcollaborationName{The Friends of AASTeX Collaboration}

\begin{document}


\title{Determining the core radio luminosity function of radio AGNs via copula}


\author{Zunli Yuan\altaffilmark{1,2,3}, Jiancheng Wang\altaffilmark{1,2,3}, D. M. Worrall\altaffilmark{4}, Bin-Bin Zhang\altaffilmark{5,6}, Jirong Mao\altaffilmark{1,2,3}}
\affil{yuanzunli@ynao.ac.cn}







\altaffiltext{1}{Yunnan Observatories, Chinese Academy of Sciences, Kunming 650216, P. R. China}
\altaffiltext{2}{Key Laboratory for the Structure and Evolution of Celestial Objects, Chinese Academy of Sciences, Kunming 650216, P. R. China}
\altaffiltext{3}{Center for Astronomical Mega-Science, Chinese Academy of Sciences, Beijing 100012, P. R. China}
\altaffiltext{4}{HH Wills Physics Laboratory, University of Bristol, Tyndall Avenue, Bristol BS8 1TL, UK}
\altaffiltext{5}{School of Astronomy and Space Science, Nanjing University, Nanjing 210093, P. R. China}
\altaffiltext{6}{Key Laboratory of Modern Astronomy and Astrophysics (Nanjing University), Ministry of Education, P. R. China}

\begin{abstract}
The radio luminosity functions (RLFs) of active galactic nuclei (AGNs)
are traditionally measured based on total emission, which doesn't reflect the current activity of the central black hole. The increasing interest in compact radio cores of AGNs requires determination of the RLF based on core emission (i.e., core RLF). In this work we have established a large sample (totaling 1207) of radio-loud AGNs, mainly consisting of radio galaxies (RGs) and steep-spectrum radio quasars (SSRQs). Based on the sample, we explore the relationship between core luminosity ($L_c$) and total luminosity ($L_t$) via a powerful statistical tool called ``Copula". The conditional probability distribution $p(\log L_{c} \mid \log L_{t})$ is obtained. We derive the core RLF as a convolution of $p(\log L_{c} \mid \log L_{t})$ with the total RLF which was determined by previous work. We relate the separate RG and SSRQ core RLFs via a relativistic beaming model and find that SSRQs have an average Lorentz factor of $\gamma=9.84_{-2.50}^{+3.61}$, and that most are seen within $8^{\circ} \lesssim \theta \lesssim 45^{\circ}$ of the jet axis. Compared with the total RLF which is mainly contributed by extended emission, the core RLF shows a very weak luminosity-dependent evolution, with the number density peaking around $z\thicksim 0.8$ for all luminosities. Differences between core and total RLFs can be explained in a framework involving a combination of density and luminosity evolutions where the cores have significantly weaker luminosity evolution than the extended emission.

\end{abstract}





\section{Introduction}
\label{Intro}

Observations have suggested that radio-loud active galactic nuclei (AGNs) play an important role in feedback, and thus have
a significant impact on galaxy evolution \citep[e.g.,][]{2006MNRAS.370..645B,2006MNRAS.365...11C,2012ARAA..50..455F,2013MNRAS.436.1084M}.
This type of AGN, which at high powers includes radio galaxies (RGs) and quasars, displays double lobes connected to a ``core'' by jets
on scales of $\thicksim$100 kpc. In unification schemes, quasars are the beamed end-on counterparts of RGs. A RG can be generically described
by a three-component structure of core, jets and lobes. The core, which is traditionally defined as a component unresolved on
arcsecond scales and with a flat radio spectrum \citep[e.g.,][]{1998MNRAS.296..445H,2008MNRAS.390..595M}, is one of the most
important structures in radio-loud AGN as it marks where the active nucleus propels energy
and matter to extended lobes via jets. The standard interpretation that the core is the optically thick base of the
jet \citep[e.g.,][]{1979ApJ...232...34B}, has been confirmed by VLBI maps \citep[e.g.,][]{2011arXiv1101.0837A}.

The radio core emission is generally thought to be self-absorbed nonthermal synchrotron emission originating in the inner
jet \citep[e.g.,][]{2002AJ....123.1334V,2004AA...425..825K,2018ApJ...861..129K}. It is directly associated with processes in the central engine, and
related to accretion and triggering of the supermassive black hole (SMBH). At low radio frequencies, the core is often only about 0.001 times the
flux density of the total source. The core and jets are affected by relativistic beaming that causes orientation dependencies.
The lobes, which display extended structures and are composed of old plasma, dominate the low-frequency emission of the source and
are not affected by relativistic effects but do not relate directly to current processes in the central engine.

The radio luminosity function (RLF) is an important statistical tool used to study the evolution of radio sources. Up to now most research
on the RLFs of radio-loud AGN has been based on their total radio emission \citep[i.e., total RLF, e.g.,][]{1990MNRAS.247...19D,2001MNRAS.322..536W,2011MNRAS.413.1054M,2016ApJ...820...65Y}.
In principle, we can also determine RLFs based on core emission (i.e., core RLFs), and can expect that the core RLF would be
more closely associated with the fundamental physical processes creating and maintaining jets than the total RLF which is strongly affected by time-averaged properties and the environment.

The motivation of this work is based on the importance of the core RLF. It can be important at least in the following aspects. Firstly, the
shape and evolution of the core RLF would provide more rigorous constraints on the nature of the instantaneous radio activity in massive galaxies.
Since core radio activity arises within a very small ($<$1 pc) region \citep[e.g.,][]{1995MNRAS.276.1373S}, the difference between
radio loud and radio weak AGN is established already on the parsec scale \citep{1995AA...293..665F}. Secondly, the core RLF will help us to understand
the accretion process onto SMBHs more directly than that for the total RLF: the core's radiation is closely
linked with the property of the SMBH while environmental effects play an important role for the large-scale lobes.
For example, the observed relation between BH mass, radio and X-ray luminosity (known as the Fundamental Plane of active BHs) that
defines the physical state of low kinetic mode objects \citep[see][]{2008MNRAS.388.1011M} is based on the observed (5 GHz) radio core emission \citep[e.g.,][]{2009MNRAS.396.1929H},
not on the extended one. Thirdly, galaxies have weak radio emission on extended scales that is unrelated to the AGN-related emission
(i.e., starburst related instead), and total RLFs run into the problem that they start picking up such objects at
low luminosities and so are no longer measuring AGN characteristics. That is less of a problem for a core RLF.
Fourthly, the increasing interest in compact radio cores with the forthcoming advent of the square kilometer array (SKA) requires determination of the core RLF.
The presence of a compact radio core in the nuclei of galaxies is usually believed to be a clear sign of BH activity \citep[e.g.,][]{2018MNRAS.476.3478B}.
In view of this, \citet{2004NewAR..48.1157F} argued that the radio emission from compact cores can be used effectively for large radio surveys with the SKA, and these cores can be used
to study the evolution of BHs throughout the universe and even to detect the very first generation of BHs.

Interest in the cores of RGs is reflected in studies at radio
  frequencies $\gtrsim 10$ GHz
  \citep[e.g.,][]{2013MNRAS.429.2080W,2014MNRAS.438..796S,2015MNRAS.453.4244W}
  based on, for example, the Tenth Cambridge (10 C) Survey
  \citep{2011MNRAS.415.2699A} and the Australia Telescope 20 GHz
  (AT20G) survey \citep{2010MNRAS.402.2403M}. For high-frequency
  selected sources, the radio emission arises mainly from the core
  \citep[e.g.,][]{2006MNRAS.371..898S}, and many sources lack extended
  radio emission and are analogous to FR\,0s \citep[e.g.,][]{2015A&A...576A..38B}.
These recent studies have suggested that the radio core is a
  key component to understanding the faint source population at high-frequency \citep[also see][]{2017MNRAS.471..908W}.

Up to now, observed data on core flux densities have been abundant, but establishing a complete core sample with a good control over the selection function is still rather
difficult. On one hand, at low frequencies radio surveys of AGN are selected based on total emission but not on core emission. Obviously, completeness in total flux is not
the same thing as completeness in core flux. On the other hand, at high frequencies a flat-spectrum core is dominant, and so flux-limited complete samples at high frequencies
are biased towards quasars and sources with bright beamed core emission. Therefore, the relativistic beaming effect brings further difficulties to the estimation of
core RLF. Due to the above factors, a comprehensive and reliable description of the core RLF is still absent.

To estimate the core RLF, some more sophisticated statistical approach should be adopted, in which the problem of sample completeness as well as the relativistic
beaming of core emission are taken into account. In regard to beaming, our plan is to use a steep-radio-spectrum source sample only, which will be discussed in
section \ref{sample}. On the issue of sample completeness problem, coincidentally, the difficulty in estimating the core RLF is very similar to that in determining the black hole
mass functions (BHMFs). The BHMF is derived by applying the existing relations between $M_{BH}$ and host galaxy properties to galaxy luminosity or velocity functions \citep[e.g.,][]{2004MNRAS.351..169M}. Similarly, we can derive the core RLF by applying a relation between core and total radio luminosities to the total RLF, which is well determined.
To give a mathematically rigorous description of the core-total relation, we resort to a special statistical tool called `Copula', which is developed by modern statistics to describe the dependence between random variables. In recent years, copula has been widely used in various areas such as finance and hydrology, but its application in astronomy or astrophysics is limited \citep{2009MNRAS.400..219B,2009AJ....137..329J,2009MNRAS.393.1370K,2010ApJ...708L...9S,2010MNRAS.406.1830T,2017MNRAS.471.2771K}.

The paper is organized as follows. Section \ref{sample} describes the properties of the sample. In Section \ref{data_analysis}, the core to total radio luminosity correlation is analyzed. Section \ref{copula} introduces the concept of copula, and presents the correlation described by copula. The core RLF is derived in Section \ref{coreRLF}. Section \ref{discussion} discuss the difference between core and total RLFs. The main results of the work are summarized in Section \ref{sum}. Throughout the paper, we adopt a Lambda Cold Dark Matter cosmology with the parameters $\Omega_{m}$ = 0.27,  $\Omega_{\Lambda}$ = 0.73, and $H_{0}$ = 71 km s$^{-1}$ Mpc$^{-1}$.

\section{The sample}
\label{sample}
Radio-loud quasars are traditionally classified in two main categories: steep spectrum (SSRQs $\alpha >0.5$, assuming $S_{\nu} \propto \nu^{-\alpha}$) and flat spectrum (FSRQs, $\alpha < 0.5$). According to unification schemes \citep[e.g.,][]{1993ARAA..31..473A,1995PASP..107..803U}, the appearance of the steep/flat-spectrum dichotomy depends primarily on axis orientation relative to the observer, while intrinsic properties are similar. Steep-spectrum sources include RGs and SSRQs, and are lobe-dominated and inclined at larger angles to the line of sight compared with their flat-spectrum counterparts. Due to the relatively larger viewing angles, the radio cores in steep-spectrum sources are much less affected by Doppler-boosting compared with those in flat-spectrum sources. Therefore, we will use a steep-spectrum source sample only to determine the core RLF. Obviously,
this choice will lead to missing many cores of flat-spectrum sources but the bias can be quantified as long
as the unification scheme of AGNs is true and the inclination angles of radio sources are randomly distributed. The core RLF derived from the steep-spectrum sources
would then be different from the intrinsic core RLF only in normalization factor, but not in shape \citep[e.g.,][]{2007ApJ...667..724L}.

\subsection{The sample composition}
This work involves two samples, referred to as Samples $I$ and $II$. Sample $I$ is a complete ``coherent'' \citep[e.g.,][]{1980ApJ...235..694A} sample consisting of four subsamples with different flux limits. It was established by our previous work \citep{2012ApJ...744...84Y}, and was used to determine the total RLF by \citet[][hereafter Y17]{2017ApJ...846...78Y}. Y17's total RLF is the important base for this work. Sample $II$ is the base to explore the relationship between core and total luminosities via copula. It inherits all the sources (totaling 631) which have both total and core flux density measurements from Sample $I$. It also absorbs the 73 sources from the GRG (giant RG) sample by \citet{2004AA...421..899L}. Through an extensive literature search we collect 503 additional sources and put them into Sample $II$. The list of these 503 sources can be found in the Appendix \ref{details}. Sample $II$ thus includes 1207 radio-loud AGNs which mainly consist of RGs and SSRQs. In statistics, a simple random sample is a subset of individuals (a sample) chosen from a larger set (a population). Each individual is chosen randomly and entirely by chance \citep{Yates2008}. The data of Sample $II$ are collected from various sources. It can be treated approximately as a simple random sample.

\subsection{Sample II}
All the sources in Sample $II$ have radio core flux densities at 5 GHz, total radio flux densities at 408 MHz or 1.4 GHz, and redshifts. The source composition, and the numbers of sources for which parameters of interest are measured are shown in Table \ref{tab:num}. In the table, $S_{c5.0}$ represents the radio core flux density at 5 GHz and $z$ is redshift. $S_{t0.408}$ and $S_{t1.4}$ represent the total radio flux densities at 408 MHz and 1.4 GHz, respectively. $\alpha_{t}$ is the spectral index near 408 MHz for total emission, and $\alpha_{c}$ represents the core spectral index near 5 GHz. Note that the 73 sources from \citet{2004AA...421..899L} only have total radio flux densities measured at 1.4 GHz. We will apply spectral indices to them using a Monte Carlo method (see section \ref{MC} for detail), and then convert $S_{t1400}$ to $S_{t408}$ so that for all the sources, a monochromatic luminosity at 5 GHz for cores, and 408 MHz for total emission can be calculated. Throughout the paper, when it comes to the core and total luminosities (denoted as $L_c$ and $L_t$ respectively), we always mean the 5 GHz and 408 MHz monochromatic luminosities, respectively.

\begin{table}[!t]
\tablewidth{0pt}
\renewcommand{\arraystretch}{1.5}
\caption{Completness of the data}
\begin{center}
\begin{tabular}{lccccccc}
\hline\hline

\colhead{Ident.} & \colhead{$z$} & \colhead{$S_{c5.0}$} & \colhead{$S_{t0.408}$} & \colhead{$S_{t1.4}$} & \colhead{$\alpha_{t}$} & \colhead{$\alpha_{c}$} & \colhead{Total} \\
\hline
 RGs     &752 & 752 &682 & 70 & 682 & 388 & 752 \\
 quasars &455 & 455 &452 & 3  & 452 & 232 & 455 \\
\hline
\end{tabular}
\end{center}
\label{tab:num}
\end{table}

\section{Data analysis}
\label{data_analysis}

\subsection {Mathematical notation}
We use an italic capital letter to denote a random variable; e.g., $L_c$ is the core luminosity or its value, while \emph{$L_C$} denotes the random variable. We use the common statistical notation that an estimate of a quantity is denoted by placing a ``hat" above it; e.g., $\hat{\theta}$ is an estimate of the true value of the parameter $\theta$. We use a non-parametric method, called kernel density estimation (KDE), to estimate the probability density function (PDF) of a random variable. Let $(x_1, x_2, \cdots, x_n)$ be a univariate independent sample drawn from some distribution with an unknown density $f(x)$. The KDE of this function $f$ is given by
\begin{eqnarray}
\label{kde}
f(x)\cong \hat{f}_h(x)=\frac{1}{nh}\sum_{i=1}^nK(\frac{x-x_i}{h}),
\end{eqnarray}
where $K$ is the kernel (a non-negative function that integrates to one), and $h > 0$ is a smoothing parameter called the bandwidth. The normal kernel is often used,
which means taking $K(x)$ as the standard normal density function. The bandwidth of the kernel is a free parameter which exhibits a strong influence on the resulting estimate. We follow the method of \citet{Botev2010} to chose an optimal $h$.

\subsection {The spectral index distribution}
\label{tSID}
\begin{figure}
  \centerline{
    \includegraphics[scale=0.40,angle=0]{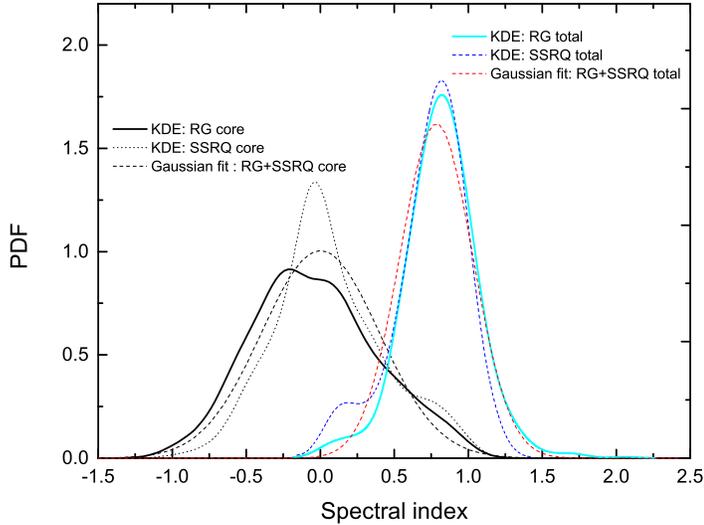}}
  \caption{\label{f_SID} Distributions of the core and total spectral indices, with the meaning of each curve being explained in section \ref{tSID}.}
\end{figure}

The distributions of spectral indices for radio core and total emission are shown in Figure \ref{f_SID}. The black thick solid and black dotted curves represent the core spectral indices of RGs and SSRQs, respectively. These curves are plotted based on the KDE. We notice that the two curves have similar mean and standard deviation. In Figure \ref{f_SID}, the black dashed curve shows the Gaussian fit for the RG+SSRQ core spectral indices. The spectral index distributions of total emission for RGs (cyan thick solid curve) and SSRQs (blue dashed curve) are even more similar to one another. The red dashed curve shows the Gaussian fit for the RG+SSRQ total spectral indices. The mean and standard deviation of Gaussian fits for the core and total spectral indices are given in Table \ref{tab:sid}.

\subsection {Dealing with the spectral index incompletness}
\label{MC}
In our RG sample, 9.3\% of the sources do not have total spectral index, and 48.3\% of the sources do not have core spectral index. For the SSRQ sample, the two percentages are 0.66 \% and 49\%, respectively. For the sources without spectral indices, we assume the spectral indices follow Gaussian distributions (with means and sigmas given in Table \ref{tab:fit}), and assign random spectral indices to them by a Monte Carlo method. We create 10000 simulated samples of the 752 RGs and 455 SSRQs, in which the sources with \emph{total} spectral index less than 0.4 \citep[e.g.,][]{2012MNRAS.422.2274C} are excluded from the analysis. The minimum spectral index criterion means statistically that all the sources entering the analysis are lobe-dominated. In the following sections, we will introduce the analysis process, which is done independently for each simulation. The final result is built as the average of the results, and its uncertainty takes into account the spread of all the Monte Carlo results \citep[also see][]{2014ApJ...780...73A}.

\begin{table}[!t]
\tablewidth{0pt}
\renewcommand{\arraystretch}{1.5}
\caption{Gaussian fits to the spectral index distribution}
\begin{center}
\begin{tabular}{lcc}
\hline\hline

\colhead{} & \colhead{RG+SSRQ core} & \colhead{RG+SSRQ total }  \\
\hline
 Mean &0.001 & 0.785  \\
 Sigma &0.397 & 0.246  \\
\hline
\end{tabular}
\end{center}
\label{tab:sid}
\end{table}

\subsection {The core-total radio luminosity correlation}
\begin{figure}
  \centerline{
    \includegraphics[scale=0.40,angle=0]{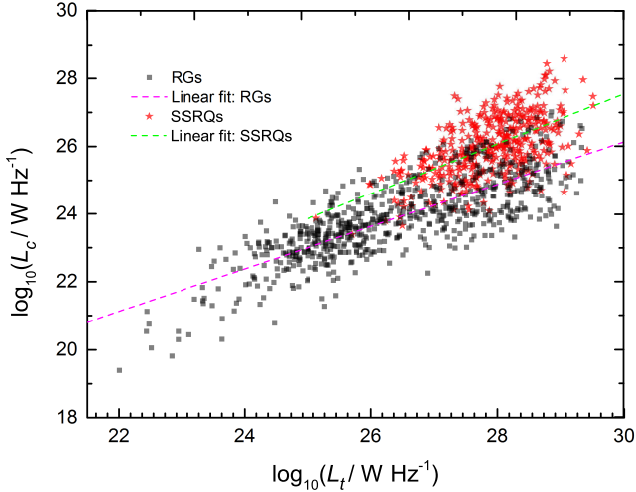}}
  \caption{\label{LcLt} Correlation of core luminosity at 5 GHz vs. total luminosity at 408 MHz. The black squares and red stars represent RGs and SSRQs, respectively. The magenta and green dashed lines show the linear fits, i.e., $\log L_c=\log L_t\times(0.63\pm0.02)+(7.34\pm0.48)$ for RGs, and $\log L_c=\log L_t\times(0.74\pm0.05)+(5.42\pm1.47)$ for SSRQs.}
\end{figure}

It is noticed that there is a correlation between the core and total
radio luminosities in radio AGNs
\citep[e.g.,][]{1988AA...199...73G,1995ApJ...448..521Z,2004AA...421..899L,2002AA...381..757L}. In
Figure \ref{LcLt}, the core luminosity versus the total luminosity for
our sample is plotted, with the RGs and SSRQs being shown as black
squares and red stars, respectively. Statistically, the core and total
radio luminosities can be regarded as random variables \emph{$L_C$}
and \emph{$L_T$}. The $L_c-L_t$ correlation means that a dependence
exists between \emph{$L_C$} and \emph{$L_T$}. However, caution
  must be taken when treating the $L_c-L_t$ correlation, because both
  $L_C$ and $L_T$ may strongly correlate with redshift and this could result in a spurious luminosity correlation \citep[e.g.,][]{1992A&A...256..399P}. The proper way of dealing with the problem is to examine the $L_c-L_t$ correlation eliminating the effect of redshift, i.e., via a partial correlation analysis \citep[e.g.,][see Appendix \ref{PCAs} for details]{2011MNRAS.413..852G,2011ApJ...733...66I}. This is performed for our Monte Carlo simulated samples. We calculate that the average partial correlation coefficients and \emph{p}-values are 0.289 and $1.002\times10^{-14}$ for RGs, and 0.232 and $3.910\times10^{-6}$ for SSRQs, respectively. Thus the partial correlation analysis suggests that the $L_c-L_t$ correlation is genuine.

Traditionally, the $L_C-L_T$ dependence was assumed to be linear in logarithmic space. For example, \citet{1995ApJ...448..521Z} found $\log L_c=\log L_t\times(0.56\pm0.04)+(9.0\pm1.0)$ for RGs. Based on high quality data of the core flux density observed with VLBI, \citet{2001ApJ...552..508G} found $\log L_c=\log L_t\times(0.62\pm0.04)+(7.6\pm1.1)$ for their RG sample. These are very similar to our result that the linear fit is $\log L_c=\log L_t\times(0.63\pm0.02)+(7.34\pm0.48)$ for RGs (the magenta dashed line in Figure \ref{LcLt}). However, from the perspective of statistics, the linear correlation does not rest on a strong mathematical foundation. In the modern field of statistics, scientists have developed a special statistical tool called `Copula' to describe the dependence between random variables. Besides the linear dependence, we can capture the nonlinear, asymmetric and tail dependence between variables by copula functions.

\section {Copula}
\label{copula}
\subsection {A brief introduction}
Briefly speaking, copulas are functions that join or ``couple'' multivariate distribution functions to their one-dimensional marginal distribution functions \citep{Nelson}. According to the Sklar's theorem, let $H$ be a joint distribution function with marginal distribution functions $F$ and $G$, if $F$ and $G$ are continuous, then there exists a unique copula $C$ such that
\begin{eqnarray}
\label{sklar}
H(x,y)=C(F(x),G(y)).
\end{eqnarray}
Let $u_n=F(x_n)$ and $v_n=G(y_n)$, $n=1,2,...N$, obviously $u_n$ and $v_n$ obey the uniform distribution on $[0,1]$. Hence a copula $C(u,v)$ can be regarded as the joint distribution of random vectors $(U,V)$ whose one-dimensional margins are uniform distributions on $[0,1]$ \citep{Nelson1999}. Concerning rigorous definition and detailed introduction of copula, we refer the interested reader to \citet{Nelson}.

As a joint distribution function, $H$ not only carries the information on the marginal distribution of each variable, but also implies the dependence properties between variables.
The main appeal of Equation (\ref{sklar}) is that by using copulas one can model the dependence structure and the marginal distributions separately. All the information on the dependence between variables is carried by the copula.
From Equation (\ref{sklar}), the joint probability density function $h(x,y)$ can be written as
\begin{eqnarray}
\label{hxy}
h(x,y)=c(F(x),G(y))f(x)g(y),
\end{eqnarray}
where $f(x)$ and $g(y)$ are the marginal PDFs, and $c(u,v)$ is given by
\begin{eqnarray}
\label{cuv}
c(u,v)=\frac{\partial^2 C(u,v)}{\partial u\partial v}
\end{eqnarray}
The conditional probability density function of $Y$ given the occurrence of the value $x$ of $X$ can be written as
\begin{eqnarray}
\label{cpdf}
f_Y(y|X=x)\equiv \frac{h(x,y)}{f(x)}= c(F(x),G(y))g(y).
\end{eqnarray}

Copulas consists of many families, of which the elliptical and Archimedean Copulas are most common. For example, the normal copula is an elliptical copula given by:
\begin{eqnarray}
\begin{aligned}
\label{Gaussian}
C_{\rho}(u,v)=& \int_{-\infty}^{\Phi^{-1}(u)} \int_{-\infty}^{\Phi^{-1}(v)} \\
&\frac{1}{2\pi\sqrt{1-\rho^2}} \exp\left[-\frac{s^2-2\rho st+t^2}{2(1-\rho^2)}\right] dsdt,
\end{aligned}
\end{eqnarray}
where $\Phi^{-1}$ is the inverse of the standard normal distribution function and $\rho$, the linear correlation coefficient, is the copula parameter.

\subsection {Copula modeling}
\label{copulamodel}
The purpose of copula modeling is to find an optimal copula function and also estimate its parameters to describe the observed data $(X_i,Y_i)$. In this work, we use the maximum likelihood estimate (MLE) method to estimate the parameters of a copula function. For some target copula with the parameter $\theta$, the likelihood function of the sample $(X_i,Y_i)$, $(i=1,2,...,n)$ is given by
\begin{eqnarray}
\label{Likeli}
L(\theta)=\prod_{i=1}^n c[F(x_i),G(y_i),\theta]f(x_i)g(y_i),
\end{eqnarray}
According to the MLE, the estimate of $\theta$ is $\hat{\theta}$=arg max $\ln L(\theta)$.
Once the parameters $\theta$ of a group of target copula functions are estimated, we will use the Akaike information criterion \citep[AIC,][]{Akaike1974} to select an optimal copula \citep[e.g.,][]{2011PhRvD..83b3501S}. The AIC is given by
\begin{eqnarray}
\label{AIC}
AIC=-2\sum_{i=1}^{n}\ln c[F(x_i),G(y_i),\theta] +2p_k
\end{eqnarray}
where $p_k$ is the number of free parameters in the copula model. We will take the copula with the smallest $AIC$ value as the optimal copula.

\begin{figure}[!h]
\centering
\includegraphics[width=1.0\columnwidth]{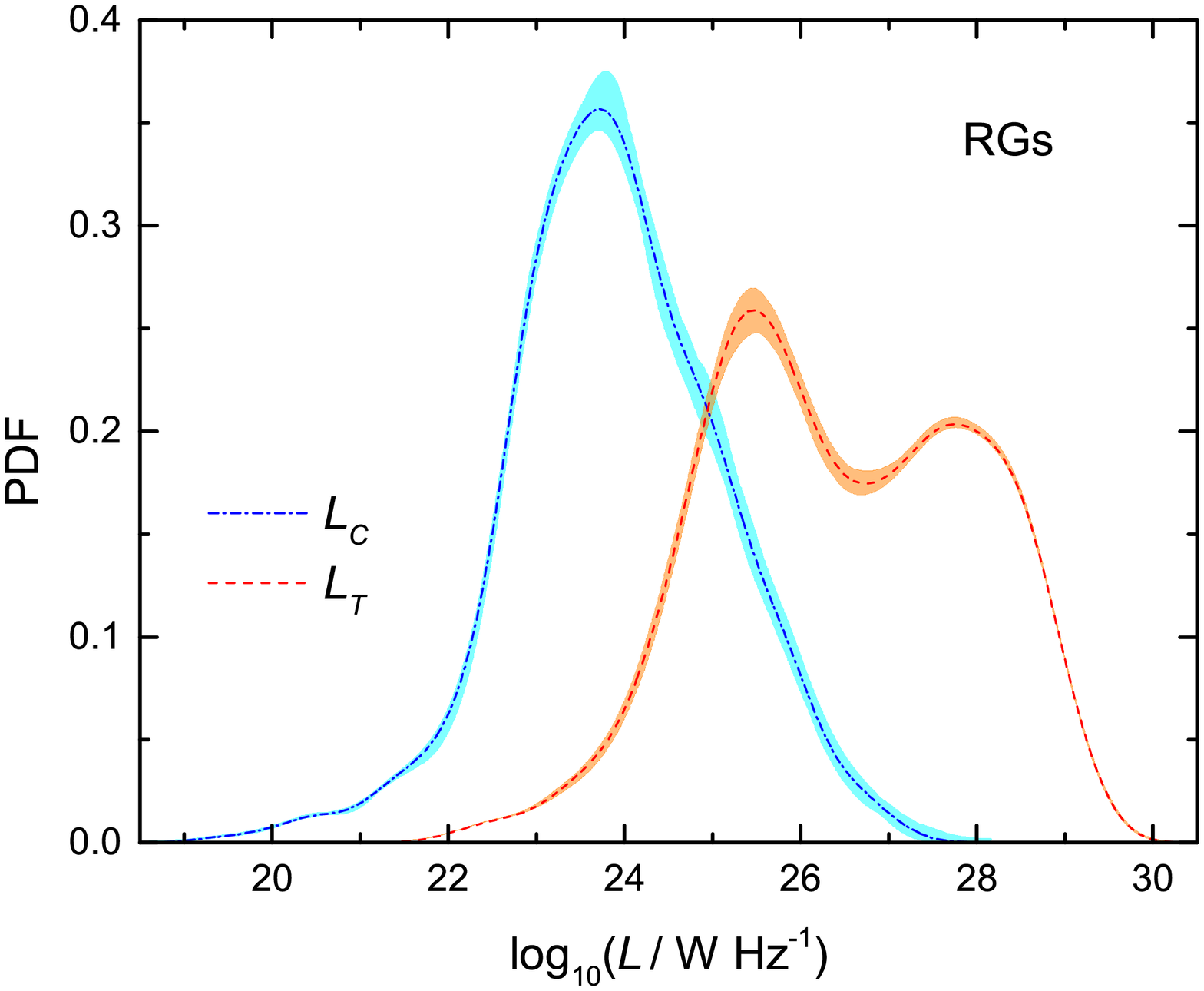}
\includegraphics[width=1.0\columnwidth]{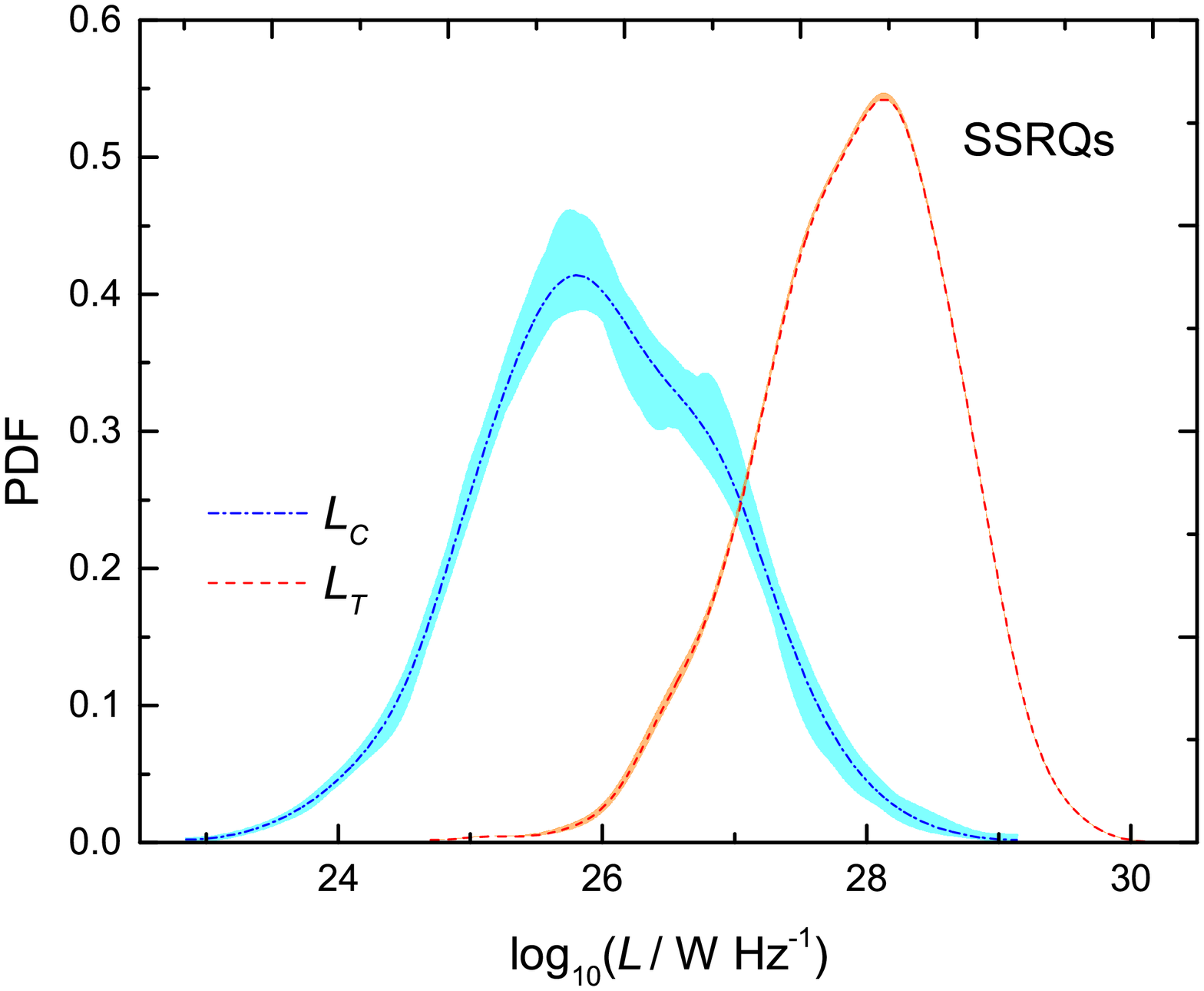}
\caption{Distribution of $L_C$ and $L_T$ for RGs and SSRQS. The light shaded areas, estimated by 10000 Monte Carlo simulations, represent the uncertainties due to the incompletness of spectral indices. The dash-dotted and dashed curves represent the average of the Monte Carlo results.}
\label{f_pdf}
\end{figure}

\subsection {Marginal PDFs}
In Equation (\ref{Likeli}), the marginal PDFs $f(x)$ and $g(y)$ need to be estimated. This can be easily realized using non-parametric estimation (e.g., KDE) or a parametric method such as MLE. Once $f(x)$ is known, $F(x)$ is simply given by
\begin{eqnarray}
\label{Fx}
F(x)=\int_{-\infty}^x f(x)dx,
\end{eqnarray}
similarly, for $G(y)$ and $g(y)$.

Take our RG sample for example, the KDE estimated marginal PDFs of
$L_T$ and $L_C$ are given in Figure \ref{f_pdf}. The red and blue
dashed curves show the KDE result from the average of the Monte Carlo
simulations. The shaded orange and cyan bands represent the
uncertainty taking into account the spread of all the Monte Carlo
simulation results. The reason for the bimodal
  shape of the $\mathrm{PDF}(L_t)$ for RGs is presumably due to a
    deficit of FRI/II boundary sources in our RG sample. It
  is well known that the FRI and II morphological classifications
  \citep{1974MNRAS.167P..31F} strongly correlate with radio power: radio sources with $L_{408~\mathrm{MHz}}\lesssim10^{25}~\mathrm{W Hz}^{-1}$ are dominated by FR Is while those with $L_{408~\mathrm{MHz}}\gtrsim10^{27}~\mathrm{W Hz}^{-1}$ are almost exclusively FR IIs \citep{1995ApJ...448..521Z}. The unimodal shape of $\mathrm{PDF}(L_c)$ for RGs indicates that the difference between radio core powers of FR Is and FR IIs is less than the difference between the extended radio powers, consistent with the study by \citet{1995ApJ...448..521Z}.

Note that for both RGs and SSRQs, the KDE estimated $\mathrm{PDF}(L_c)$ is still not smoothed enough to take as an ideal approximation of the true PDF, and this will affect the smoothness of the final core RLF. We then use a parametric method to estimate the marginal PDF $g(\log L_c)$, i.e., model it as a normal distribution
\begin{eqnarray}
\label{glc}
g(\log L_c)=\frac{1}{\sqrt{2\pi\sigma^2}}e^{-\frac{(\log Lc-\mu)^2}{2\sigma^2}}
\end{eqnarray}
where $\mu$ and $\sigma$ are free parameters to be estimated by MLE.

\subsection {Copulas for $Lc-L_t$}
We have examined 31 published copulas and applied the procedure introduced in section \ref{copulamodel} to our simulated samples to find the best two for our data. The first one is the normal copula given by Equation (\ref{Gaussian}). The second one is the number 13 Archimedean copula from \citet[][hereafter N13 copula]{Nelson} formulated as
\begin{eqnarray}
\begin{aligned}
\label{N13}
C_{\theta}(u,v)=e^{1-[(1-\ln u)^{\theta}+(1-\ln v)^{\theta}-1]^{\frac{1}{\theta}}},
\end{aligned}
\end{eqnarray}
where $\theta$ is the parameter, and $\theta\in (0,\infty)$.

In Figure \ref{AIC}, we show the distributions of the best-fit parameters of the N13 and normal copula models for our Monte Carlo samples, as well as the distributions of AIC values for the two copula modelings. The upper, and lower panels correspond to RGs and SSRQs, respectively. Table \ref{tab:AIC} summarizes the means of best-fit parameters and AIC values for the Monte Carlo samples. For both the RG and SSRQ samples, the N13 copula model has lower AIC values, and we will take it as the optimal copula.

\begin{figure}[!h]
\centering
\includegraphics[width=1.0\columnwidth]{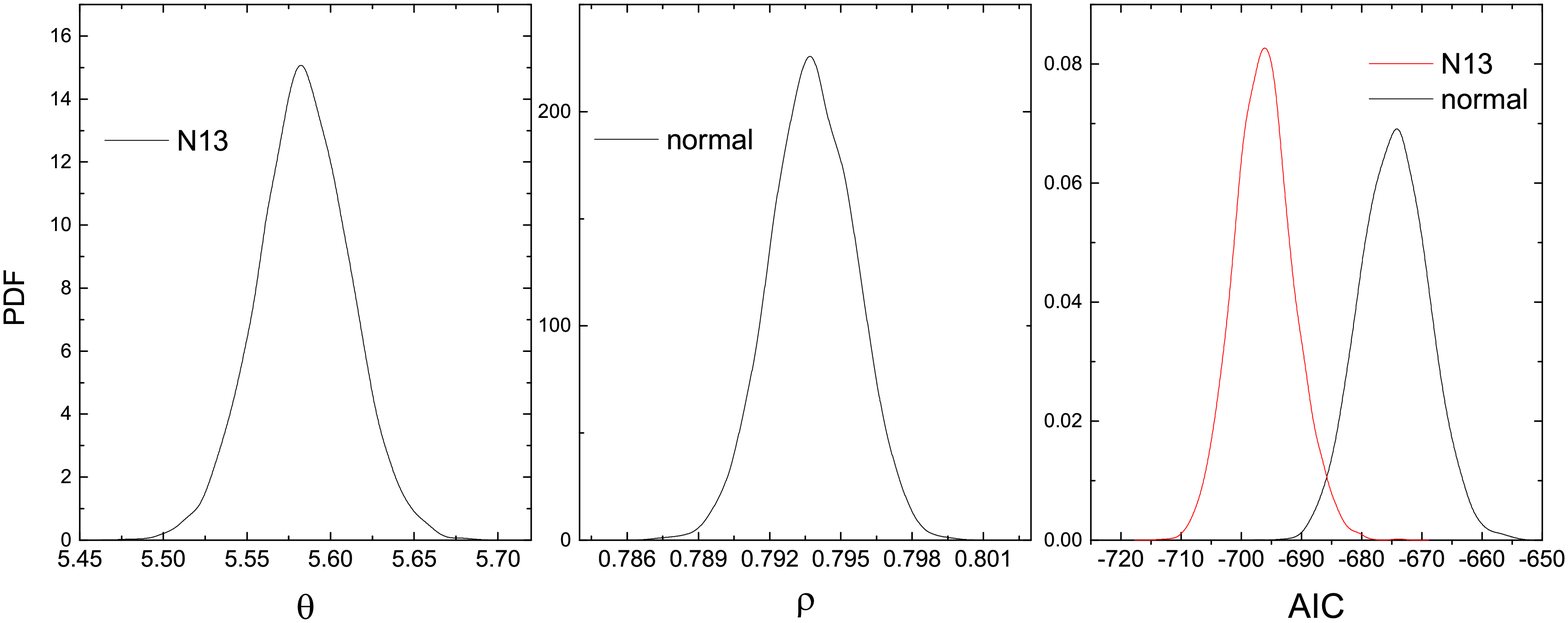}
\includegraphics[width=1.0\columnwidth]{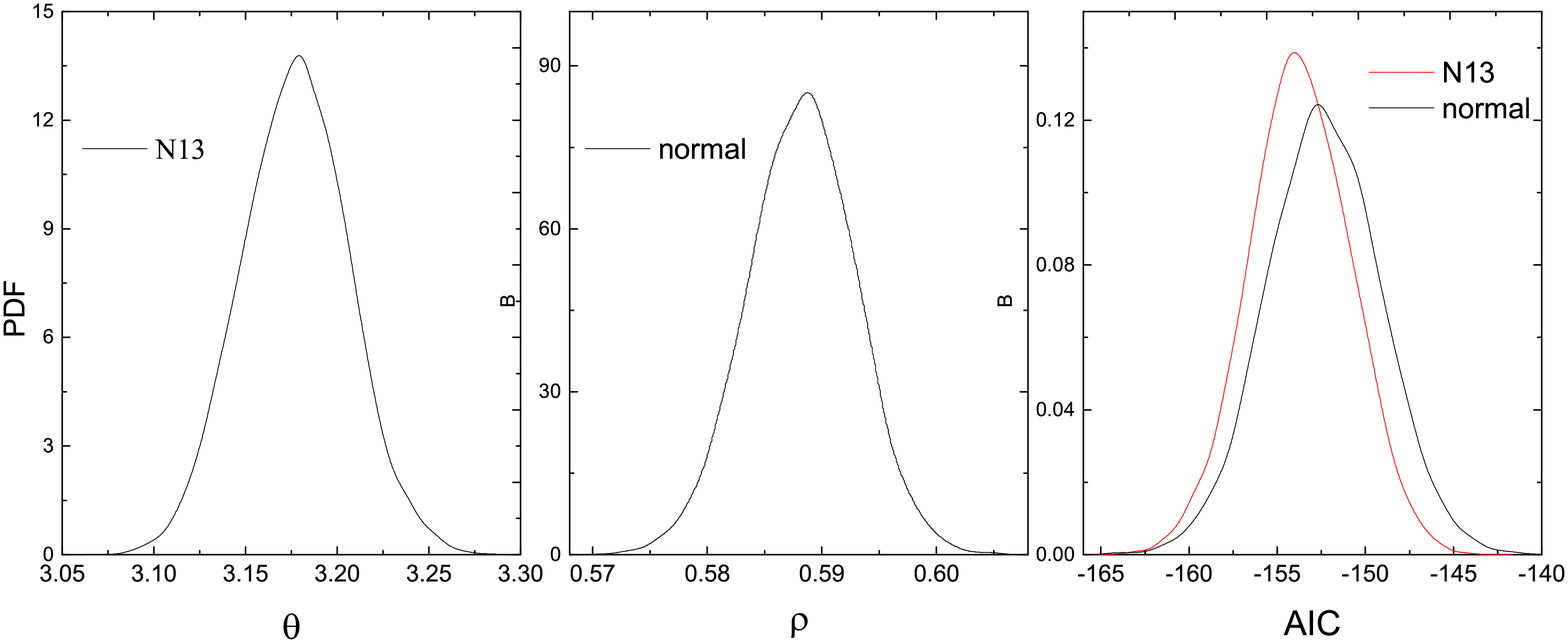}
\caption{Left and middle: Distributions of the best-fit parameters of the N13 and normal copula models for our Monte Carlo simulated samples. Right: Distributions of the AIC values for the N13 (red curve) and normal (black curve) copula modelings. The upper, and lower panels are for RGs and SSRQs, respectively.}
\label{AIC}
\end{figure}

\begin{table}[!t]
\tablewidth{0pt}
\renewcommand{\arraystretch}{1.5}
\caption{Means of the best-fit copula parameters and AIC values}
\begin{center}
\begin{tabular}{lcccc}
\hline\hline

\colhead{} & \colhead{$\hat{\theta}$} & \colhead{$\hat{\rho}$}  & \colhead{AIC N13} & \colhead{AIC normal} \\
\hline
 RGs   &5.584 & 0.794  & -696.17 & -674.55 \\
 SSRQs &3.178 & 0.588  & -153.73 & -152.35 \\
\hline
\end{tabular}
\end{center}
\label{tab:AIC}
\end{table}


\begin{table*}[!t]
\tablewidth{0pt}
\renewcommand{\arraystretch}{1.5}
\caption{Input Parameters for $\rho_t$ and Best-fit Parameters for $\rho_c$}
\begin{center}
\begin{tabular}{lcccccccc}
\hline\hline

\colhead{ }                 & \colhead{$\log_{10}\phi_1$} & \colhead{$\log_{10}L_*$}  & \colhead{$\beta$} &
\colhead{$\gamma$}          & \colhead{z$_{\rm c}$}       & \colhead{p$_{\rm 1}$}     & \colhead{p$_{\rm 2}$} & \colhead{k$_{\rm 1}$}      \\
\hline

total RLF & -4.85$_{-0.12}^{+0.13}$ & 24.68$_{-0.17}^{+0.16}$ & 0.44$_{-0.02}^{+0.02}$ & 0.31$_{-0.01}^{+ 0.01}$ & 0.86$_{-0.09}^{+0.10}$ & 0.31$_{-0.26}^{+0.22}$ & -5.92$_{-0.39}^{+0.18}$ & 4.73$_{-0.09}^{+0.16}$ \\

core RLF RG & -3.749$_{-0.008}^{+0.019}$ & 21.592$_{-0.026}^{+0.015}$ & 0.139$_{-0.007}^{+0.004}$ & 0.878$_{-0.002}^{+ 0.002}$ & 0.893$_{-0.017}^{+0.017}$ & 2.085$_{-0.077}^{+0.051}$ & -4.602$_{-0.057}^{+0.066}$ & 1.744$_{-0.050}^{+0.060}$ \\

core RLF SSRQ & -5.066$_{-0.033}^{+0.047}$ & 24.624$_{-0.073}^{+0.051}$ & 0.346$_{-0.007}^{+0.005}$ & 0.976$_{-0.009}^{+ 0.008}$ & 0.875$_{-0.021}^{+0.035}$ & 2.090$_{-0.119}^{+0.093}$ & -4.361$_{-0.106}^{+0.057}$ & 1.413$_{-0.066}^{+0.088}$ \\

\hline
\end{tabular}
\end{center}
~~~~~~~~~~~Units -- $\phi_1$: [${\rm Mpc^{-3}}$],\,\, $L_*$: [${\rm W~Hz^{-1}}$].
\label{tab:fit}
\end{table*}

\subsection {Tail dependence}
Tail dependence is an important concept in copula theory. Let $X$ and $Y$ be continuous random variables with distribution functions $F$ and $G$, respectively. The upper/lower tail dependence parameter $\lambda_U/\lambda_L$ is the limit (if it exists) of the conditional probability that $Y$ reaches extremely large/small values given that $X$ attains extremely large/small values \citep{Nelson}, i.e.
\begin{eqnarray}
\label{SED}
\lambda_U=\lim_{t\rightarrow 1^-} P[Y>G^{-1}(t)\mid X>F^{-1}(t)],
\end{eqnarray}
and
\begin{eqnarray}
\label{SED}
\lambda_L=\lim_{t\rightarrow 0^+} P[Y \leq G^{-1}(t) \mid X \leq F^{-1}(t)].
\end{eqnarray}
From the above equations, we calculate that for both the normal and N13 copulas, $\lambda_U=\lambda_L=0$. This implies that $L_T$ and $L_C$ are tail independent, meaning that when the cores reach extreme luminosities the probability that lobes also show extreme luminosities tends to zero. Physically, this can be understood as follows. The core and total (mainly contributed by extended emission) luminosities are correlated because the core and extended emission relate to the same jet kinetic power. Nevertheless, these two measurements have different timescales: extended radio luminosity is a proxy for time-averaged jet power on timescales of tens to hundreds of Myr, while core luminosity traces the instantaneous jet power \citep[see][]{2018MNRAS.478.5074S}. In addition, the lobe emission is more affected by external environment \citep[e.g.,][]{2004NewAR..48.1157F}, such as the density of intergalactic medium (IGM). Both the timescale and environment factors weaken the connection between core and extended radio luminosities. When one of them reaches extremely large/small values, the other does not response in time. Examples can be found in recurrent AGNs, as evidence is growing that AGN activity could be episodic \citep[e.g.,][]{2011MNRAS.410..484B,2009BASI...37...63S,2016ApJS..226...17L}. During the phase of inactivity, sources may lack certain features, such as radio cores or well-defined jets that are produced by continuing activity, while the radio lobes remain to undergo a period of fading before they disappear completely \citep{2009AA...506L..33M}. During the phase of reactivation, very faint fossil radio lobes remaining from an earlier active epoch can be observed, along with newly restarting jets and cores \citep{2011AA...526A.148M}. In these two situations, we can observe extremely low-luminosity cores or lobes.

\section {Determining the core RLF}
\label{coreRLF}

The RLF is defined as the number of sources per comoving volume with luminosities in the range $\log L,\log L+d\log L$
\begin{eqnarray}
\rho(z,L)=\frac{d^{2}N}{dVd\log L}.
\end{eqnarray}
We denote the total RLF as $\rho_t(z,L_t)$, and the core RLF as $\rho_c(z,L_c)$. In a previous work (Y17), we have already determined the total RLF based on a mixture evolution scenario that takes into account both density evolution (DE) and luminosity evolution (LE). Here we adopt the Model A of Y17 as the total RLF:
\begin{eqnarray}
\label{rhototal}
\begin{aligned}
\rho_t(z,&L_t)=\\
&e_{1}(z)\phi_1 \left( \frac{L_t/e_2(z)}{L_*} \right)^{-\beta} \exp\left[-\left( \frac{L_t/e_2(z)}{L_{*}}\right)^{\gamma}\right],
\end{aligned}
\end{eqnarray}
where
\begin{eqnarray}
\label{ez1}
e_1(z)=\frac{ (1+z_c)^{p_1}+(1+z_c)^{p_2} }{  \left(\frac{1+z_c}{1+z}\right)^{p_1} + \left(\frac{1+z_c}{1+z}\right)^{p_2} },
\end{eqnarray}
and
\begin{eqnarray}
\label{ez2}
e_2(z)=(1+z)^{k_1}.
\end{eqnarray}
The parameters and their 1$\sigma$ error for $\rho_t$ are given in Table \ref{tab:fit}.

\subsection {Semi-parametric core RLF}
\label{Semi_parametric_rhoc}

Considering the existence of $L_C-L_T$ correlation, the core RLF can be derived from the total RLF. The process is similar to that used to derive the BHMF \citep{2004MNRAS.351..169M}.
The difference is that their correlation description was resorted to the linear relation with a intrinsic dispersion while we use copulas.
Consulting Equation (\ref{cpdf}), and utilizing Equation (\ref{kde}), (\ref{cuv}), (\ref{Fx}), (\ref{glc}) and (\ref{N13}), the conditional PDF of $L_C$ given $L_T=L_t$ can be calculated as
\begin{eqnarray}
\label{cpd}
p(\log L_{c}|\log L_{t})=c[F(\log L_{t}),G(\log L_{t})]g(\log L_{c})
\end{eqnarray}
We then define $\rho_{t}(z,L_{t})d\log L_t$ as the number of sources per unit comoving volume at the redshift z, in the luminosity range of $\log L_t$, $\log L_t + d\log L_t$. $p(\log L_{c} \mid \log L_{t})d\log L_c$ is the probability that $L_c$ is in the range of $\log L_c$, $\log L_c + d\log L_c$ for a given $\log L_t$. Thus the number of sources with $L_c$, $L_t$ in the ranges of $\log L_c$, $\log L_c + d\log L_c$ and $\log L_t$, $\log L_t + d\log L_t$ at a redshift of z is
\begin{eqnarray}
&\rho(z,L_c,L_t)d\log L_cd\log L_t=  p(\log L_{c} \mid \log L_{t})d\log L_c\nonumber \\
&\times  \rho_{t}(z,L_{t})d\log L_t
\end{eqnarray}
Finally, The core RLF $\rho_c(z,L_c)$ is the convolution of $\rho_{t}(z,L_{t})$ and $p(\log L_{c} \mid \log L_{t})$:
\begin{eqnarray}
\label{eqn:core RLF}
\rho_c(z,L_c)=\int p(\log L_{c} \mid \log L_{t})\rho_{t}(z,L_{t})d\log L_t.
\end{eqnarray}
where the limits of integration are $\log L_{t,\mathrm{min}}=19$ and $\log L_{t,\mathrm{max}}=30$, roughly corresponding to the $L_t$ range for the RG sample.

\begin{figure}[!h]
\centering
\includegraphics[width=1.05\columnwidth]{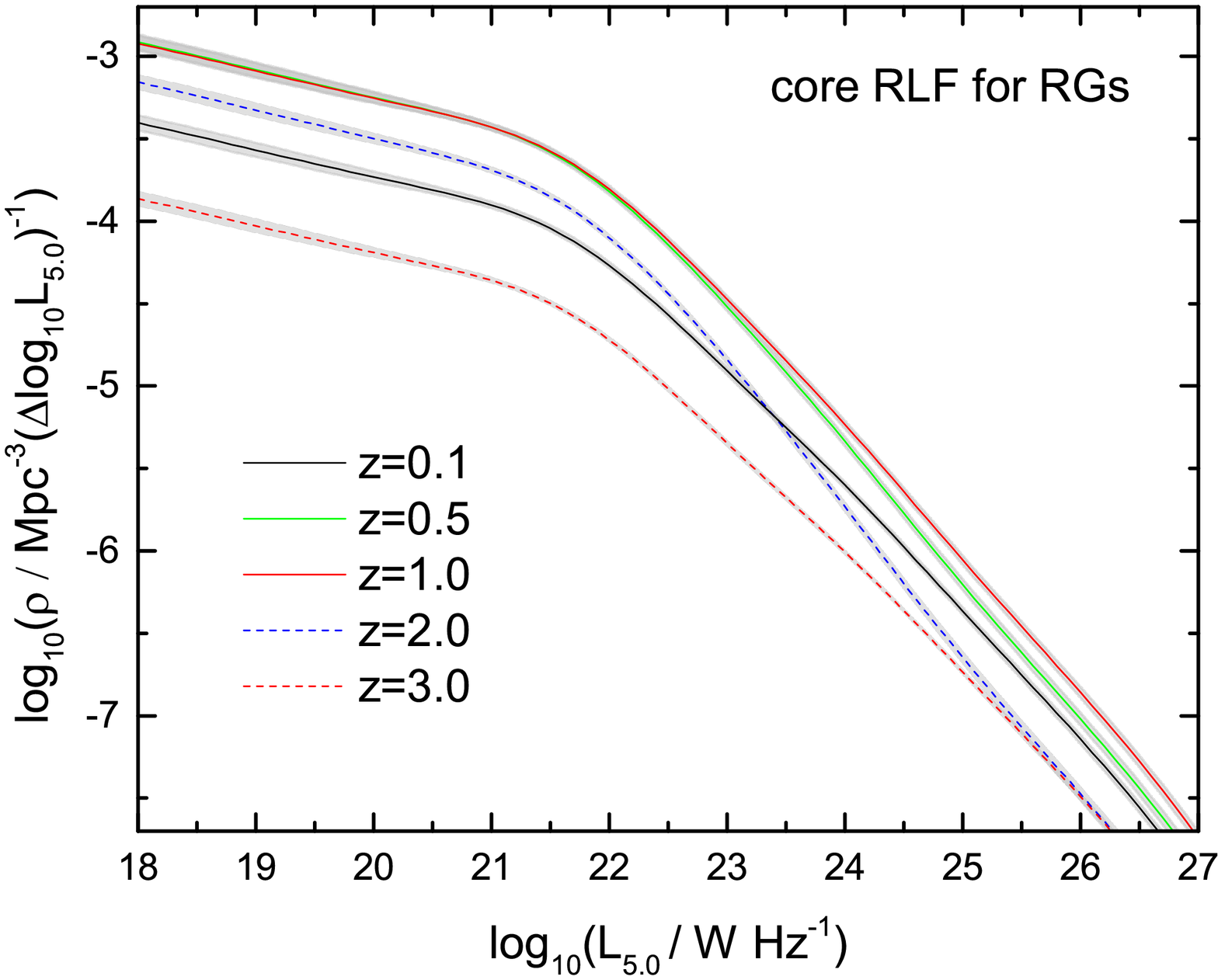}
\includegraphics[width=1.05\columnwidth]{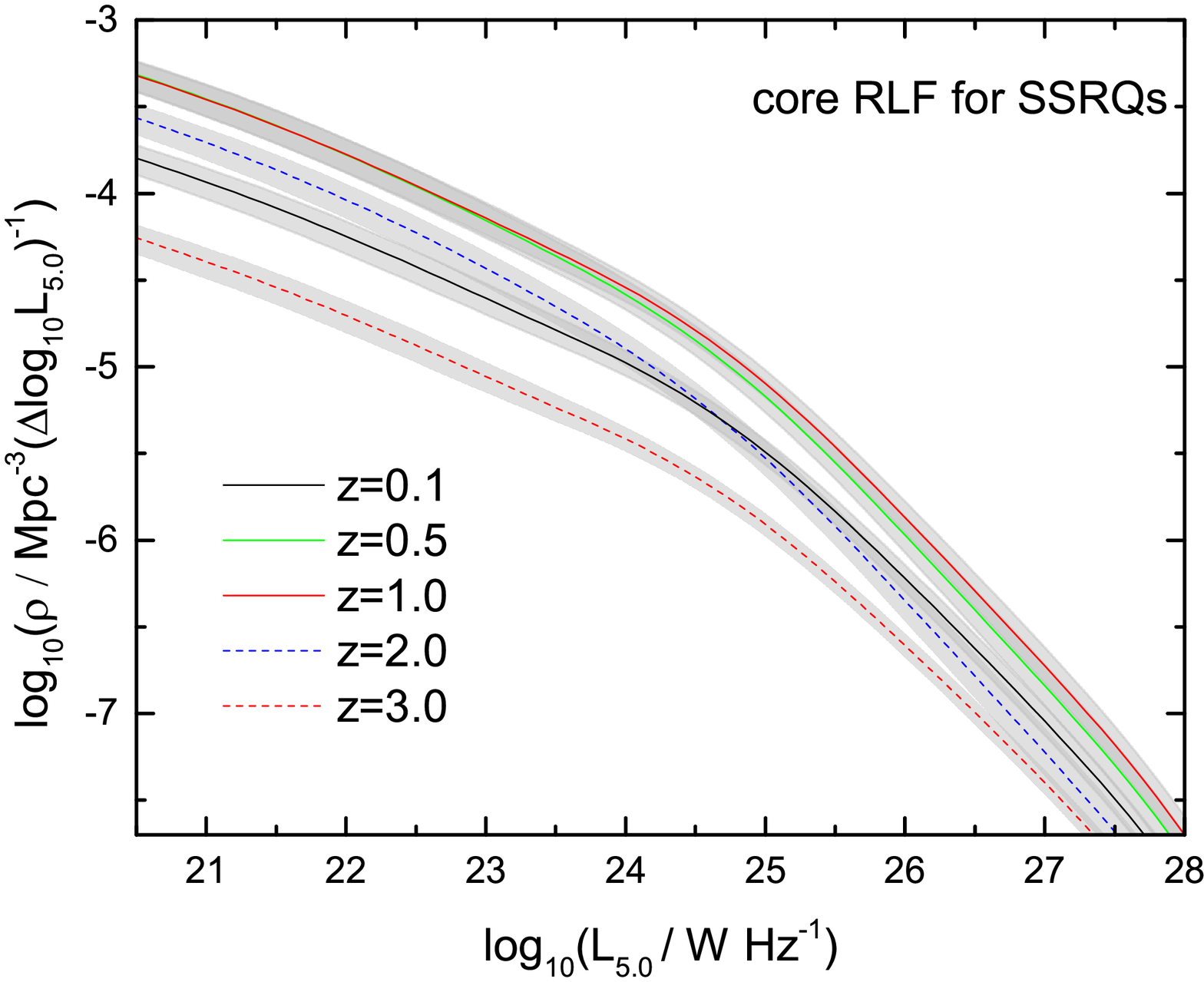}
\caption{Core RLFs derived by Equation (\ref{eqn:core RLF}) for RGs and SSRQs at $z=0.1, 0.5, 1.0, 2.0$, and $3.0$ (black, green, and red solid lines; blue and red dashed lines respectively). The gray bands, estimated by 10000 Monte Carlo simulations, represent the uncertainties due to the incompletness of spectral indices.}
\label{coreRLF1}
\end{figure}

\begin{figure}[!h]
\centering
\includegraphics[width=1.05\columnwidth]{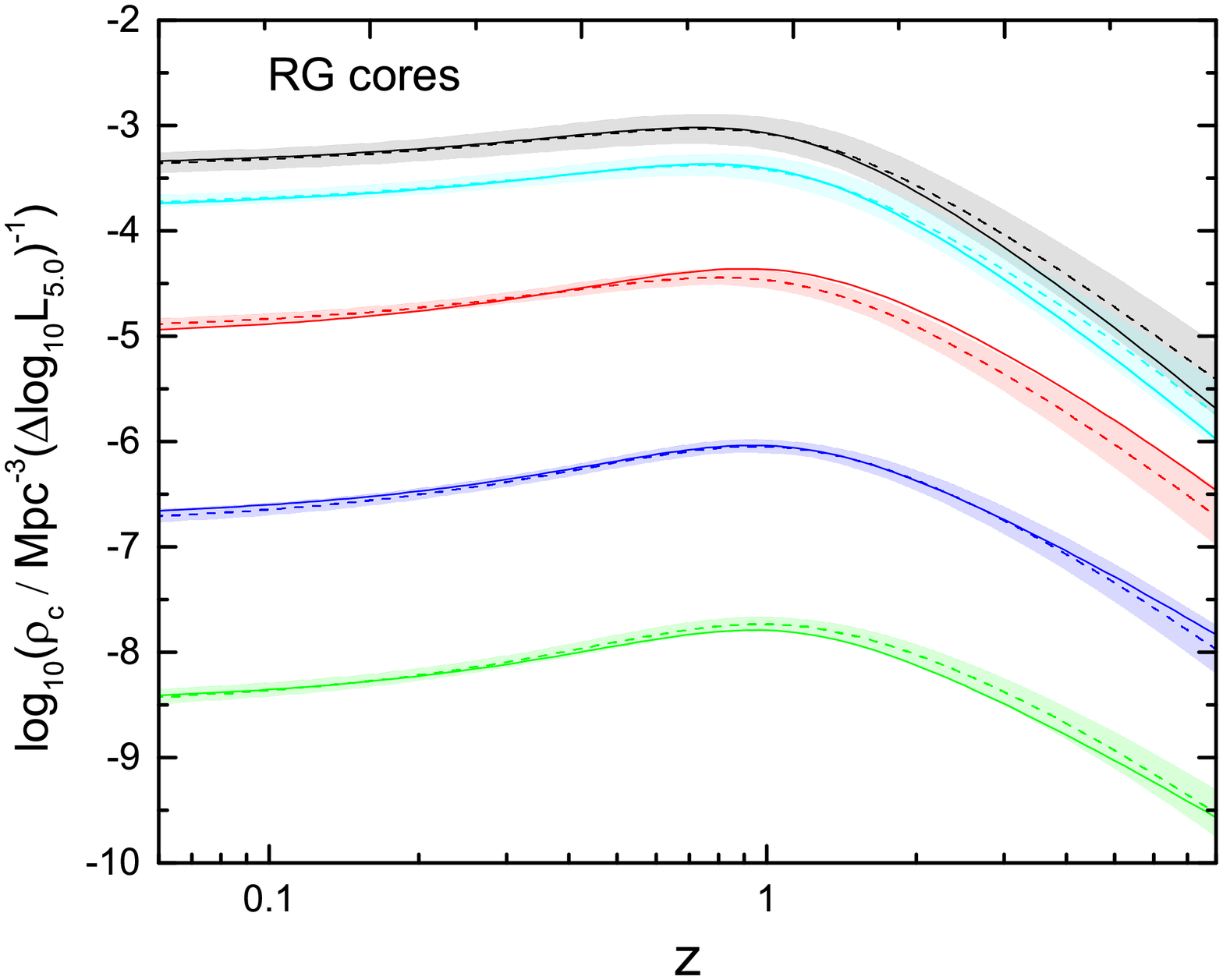}
\includegraphics[width=1.05\columnwidth]{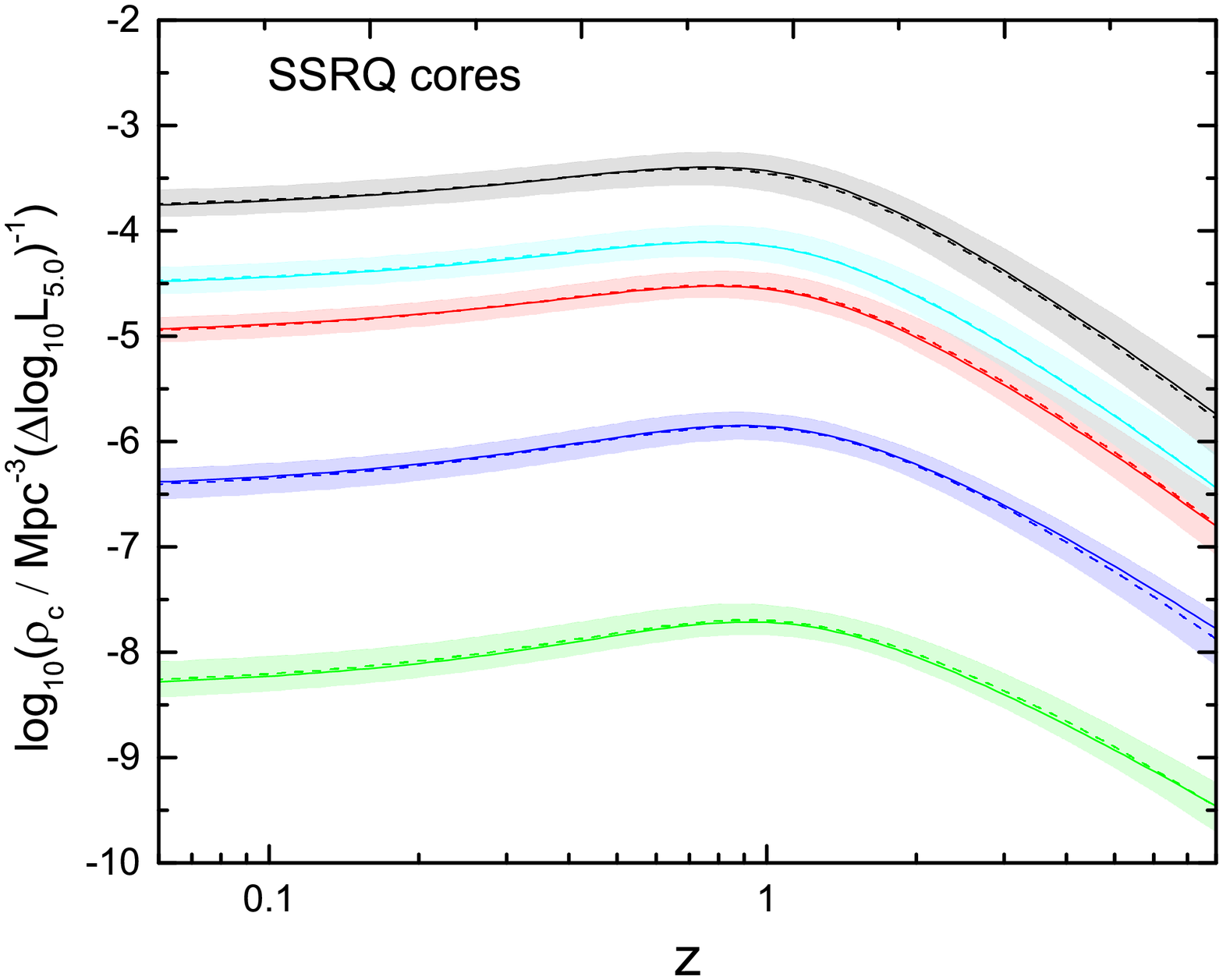}
\caption{Space densities as a function of redshift for RG and SSRQ cores. For RGs, the black, cyan, red, blue and green dashed lines show the core RLFs at $\log_{10}L_{5.0\mathrm{GHz}}$=19, 21, 23, 25 and 27, respectively. For SSRQs, these lines represent the core RLF at $\log_{10}L_{5.0\mathrm{GHz}}$=21, 23, 24, 26 and 28, respectively. The light shaded areas take into account the uncertainties due to the incompletness of spectral indices, as well as the 1 $\sigma$ error propagated from the total RLF by Y17. The solid curves represent the best-fit mixture evolution model of Section \ref{parametric_rhoc}.}
\label{RLFz}
\end{figure}

By measuring the $L_C-L_T$ correlations and corresponding copulas for the RG and SSRQ core samples separately, the core RLFs for the two populations are derived by Equation (\ref{eqn:core RLF}). Figure \ref{coreRLF1} shows the core RLFs at $z=0.1, 0.5, 1.0, 2.0$, and $3.0$ (black, green, and red solid lines; blue and red dashed lines respectively). The gray bands, estimated by 10000 Monte Carlo simulations, represent the uncertainties due to the incompletness of spectral indices. Inspection of Figure \ref{coreRLF1} suggests that the shape and evolution of the core RLFs for RGs and SSRQs are very similar. The main difference is that SSRQs have higher characteristic luminosity. This is not surprising and can be explained due to beaming. In Figure \ref{RLFz}, we show the core RLF of RGs changing with redshift at various luminosities. The black, cyan, red, blue and green dashed lines show the core RLFs at $\log_{10}L_{5.0\mathrm{GHz}}$=19, 21, 23, 25 and 27 respectively. The light shaded areas take into account not only the uncertainties due to the incompletness of spectral indices, but also the 1 $\sigma$ error propagated from the total RLF by Y17.

\subsection {Parametric core RLF}
\label{parametric_rhoc}
The core RLF given in Equation (\ref{eqn:core RLF}) is a semi-parametric function. It is not like the general luminosity functions which are obviously seen in density or/and luminosity evolutions. We use a mixture evolution model similar to that for $\rho_t$ to describe the core RLF. The only difference is replacing the modified Schechter function in Equation (\ref{rhototal}) with a double power law form:
\begin{eqnarray}
\label{rhoc}
\begin{aligned}
\rho_c(z,&L_c)=\\
&e_{1}(z)\phi_1 \left[ \left( \frac{L_c/e_2(z)}{L_*} \right)^{\beta} + \left( \frac{L_c/e_2(z)}{L_{*}}\right)^{\gamma}\right]^{-1},
\end{aligned}
\end{eqnarray}
where $e_{1}(z)$ and $e_{2}(z)$ are also given in Equations (\ref{ez1}) and (\ref{ez2}), respectively. To determine the best-fit parameters in Equation (\ref{rhoc}), we use a Bayesian Monte Carlo fitting engine (McFit) developed by \citet{2016ApJ...816...72Z}. Firstly, we take a group of uniformly-spaced points $(z_i,L_i)_{i=1,...,i=N}$ in the $\log L-\log z$ space. For each point, we calculate its $f_{data~i}$ and $\sigma_{data~i}$ by Equation (\ref{eqn:core RLF}), and $f_{mod~i}$ by Equation (\ref{rhoc}). Note that $\sigma_{data~i}$ takes into account the uncertainties due to the incompletness of spectral indices, as well as the 1 $\sigma$ error propagated from the total RLF parameters. Then the $\chi^2$ is evaluated as
\begin{eqnarray}
\label{chi2}
\chi^2=\sum_{i=1}^{N}(\frac{f_{data~i}-f_{mod~i}}{\sigma_{data~i}})^2,
\end{eqnarray}
which is related to the likelihood function by $\chi^2=-2$ln(likelihood). Based on the form of $\chi^2$, the McFit engine obtains the best-fit parameters shown in Table \ref{tab:fit}. The best-fit core RLFs are shown as solid curves in Figure \ref{RLFz}. We find that the mixture evolution model fits well the core RLFs.

\subsection {Intrinsic core RLF}
\label{Intrinsic}

\citet{1992ApJ...387..449P} estimated that SSRQs have their radio axes within $14^{\circ} \lesssim \theta \lesssim 40^{\circ}$, and high-luminosity RGs are in the range $\theta \gtrsim 40^{\circ}$. Therefore beaming is important for the cores of SSRQs, while can be neglected for the cores of RGs. Thus the core RLF of RGs is close to the intrinsic core RLF. Considering that the total RLF $\rho_t$ in Equation (\ref{eqn:core RLF}) is measured based on steep-spectrum radio sources (Y17), we estimate
\begin{eqnarray}
\label{dimauro}
\rho_{c}^{intrinsic}(z,L_c)=\kappa\rho_{c}^{RG}(z,L_c),
\end{eqnarray}
where the value of $\kappa$ should be equal to the ratio of the total number of steep- and flat-spectrum radio sources to the total number of steep-spectrum radio sources in the universe. Assuming that the steep- and flat-spectrum radio sources are divided by the critical viewing angle of $14^{\circ}$, we have $\kappa\approx 1/\cos 14^{\circ}=1.0306$. $\kappa$ is very close to one, suggesting that the RG core RLF can be regarded as the intrinsic one.

\begin{figure*}[!htb]
\includegraphics[height=7cm,width=11cm]{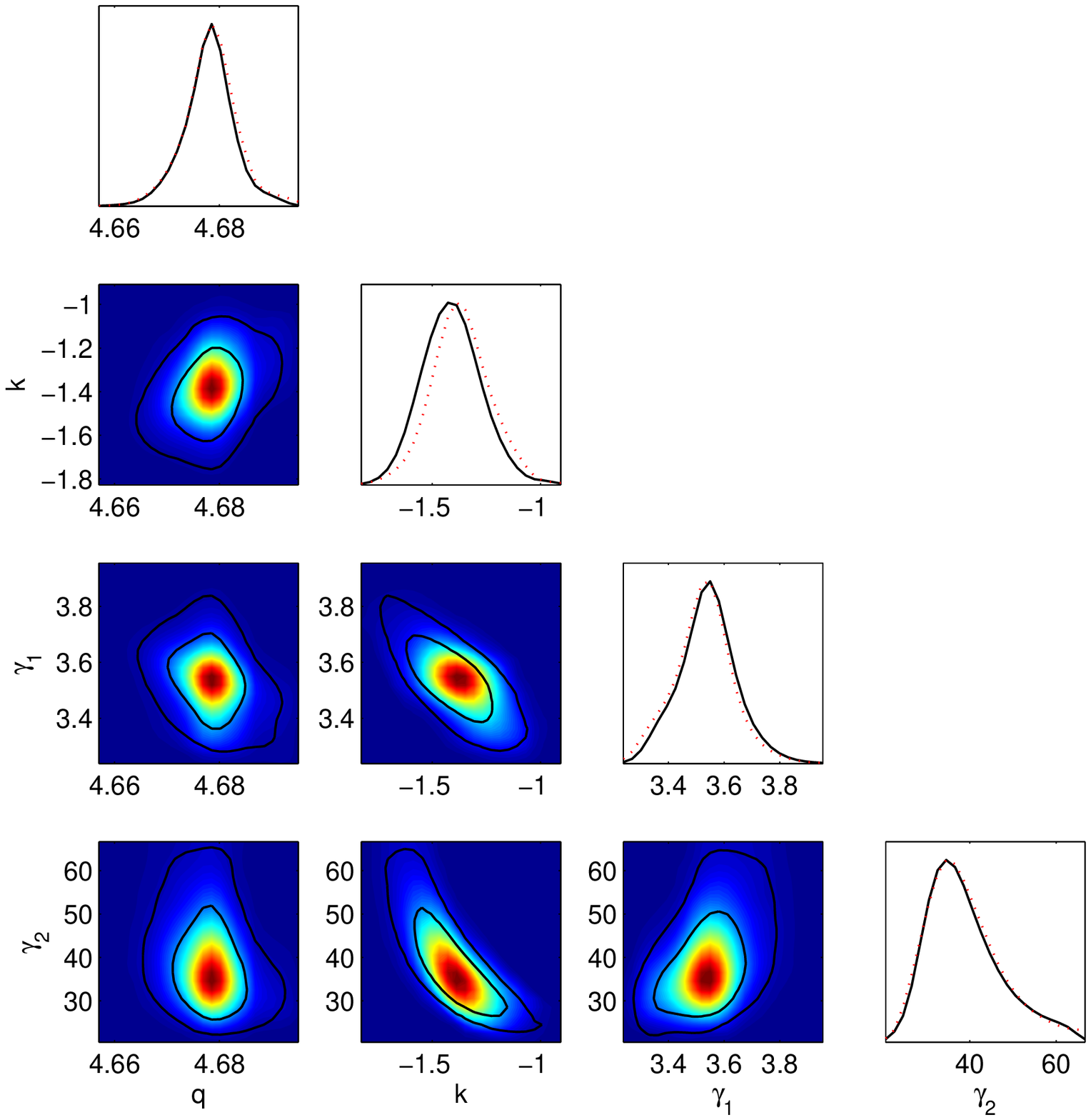}
\\[18pt]
\includegraphics[height=5cm,width=6cm]{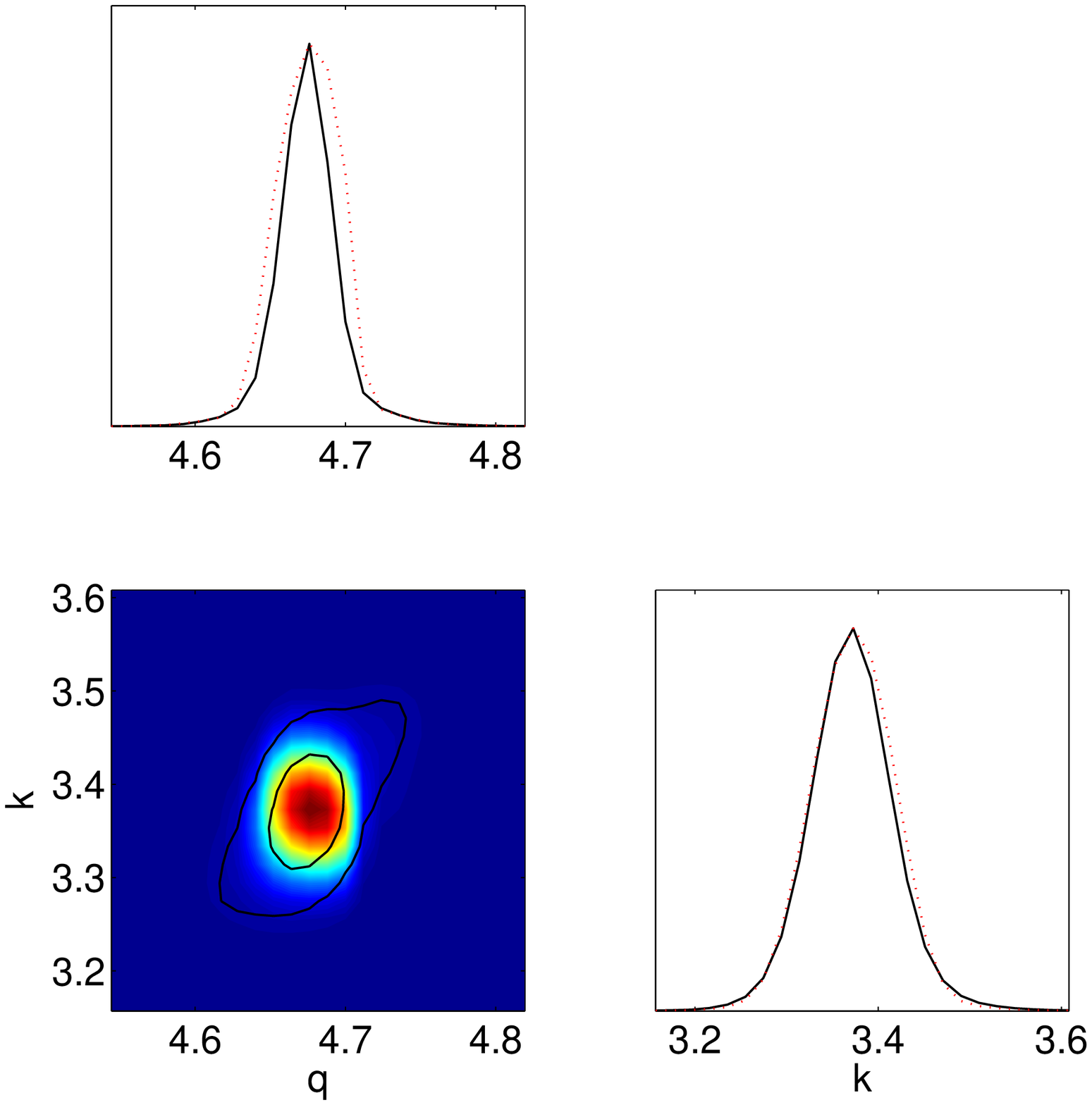}  \qquad
\includegraphics[height=7cm,width=11cm]{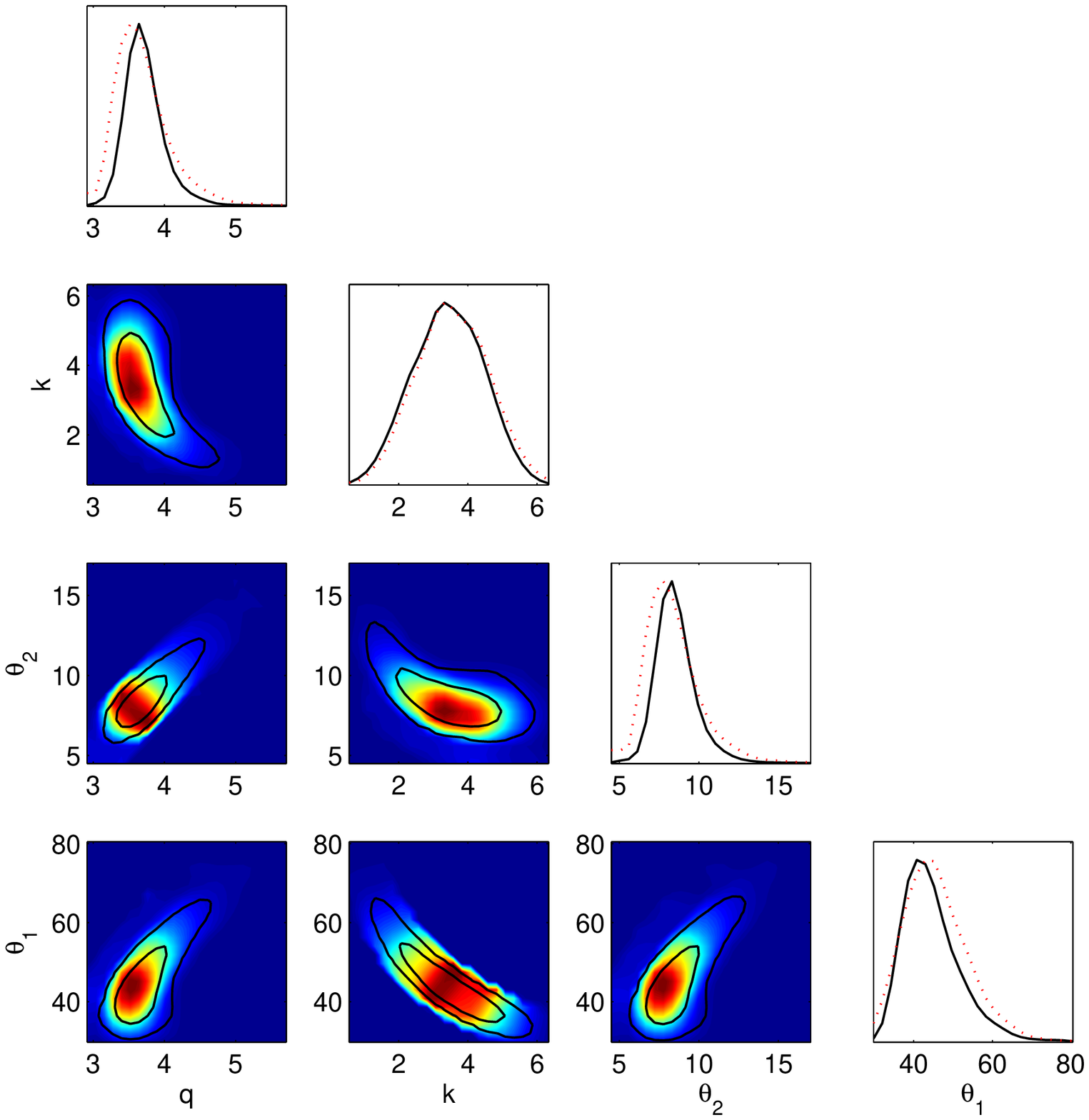}
\caption{Posterior probability distributions and 2D confidence contours of parameters in the beaming models. The red dash dot curves are the mean likelihoods of MCMC samples and the black solid curves are the marginalized probabilities. The contours are for 1 and 2 $\sigma$ levels. The upper, lower left, and lower right panels correspond to model 1, 2 and 3 respectively.}
\label{fig_mcmc}
\end{figure*}

The cores of SSRQs are expected to be the Doppler beamed counterparts of RG cores. In principle, the core RLF of SSRQs can be derived from the core RLF of RGs by considering beaming effect.
For a RG core with luminosity of $L_{c}$, after beaming it will be observed as a quasar core with luminosity of $\mathcal{L}_c$,
\begin{eqnarray}
\label{delta_p}
\mathcal{L}_{c}=L_{c} \delta^{q},
\end{eqnarray}
with $q=2+\alpha$ for a continuous jet and $q=3+\alpha$ for a moving, isotropic source \citep{1995PASP..107..803U}. Other values of $q$ are also possible, e.g., \citet{2012ApJ...751..108A} adopted a value of $q=4$ that applies to the case of jet emission from a relativistic blob radiating isotropically in the fluid frame. In Equation (\ref{delta_p}), $\delta$ is the kinematic Doppler factor defined as
\begin{eqnarray}
\label{delta_def}
\delta=\left(\gamma-\sqrt{\gamma^2-1}\cos\theta\right)^{-1},
\end{eqnarray}
where $\gamma=1/\sqrt{1-\beta^2}$ is the Lorentz factor, $\beta$ is the bulk velocity in units of speed of light, and $\theta$ is the inclination angle. To quantify the beaming effort, we need to know the PDF $P_{\delta}(\delta)$ for $\delta$. Traditionally, the jet angles are assumed to be randomly distributed within $0^{\circ} \leq \theta \leq 90^{\circ}$. Based on this assumption, the $P_{\delta}(\delta)$ was determined by \citet{2003ApJ...599..105L}. Some later researchers \citep[e.g.,][]{2007ApJ...667..724L,2008ApJ...674..111C,2012ApJ...751..108A} follow this determination. However, for a specific population of AGNs (e.g., SSRQs), the jet angles should be (randomly) distributed within $\theta_2 \leq \theta \leq \theta_1$ but not necessarily $0^{\circ} \leq \theta \leq 90^{\circ}$. Thus the formula calculating $P_{\delta}(\delta)$ by \citet{2003ApJ...599..105L} should be modified to apply to more general conditions. Here we give the generalized formula for deriving $P_{\delta}(\delta)$ (see the Appendix \ref{DFSs}, Equation (\ref{PDF_delta}) for its detailed definition and deduction) as
\begin{eqnarray}
\label{PDFdelta}
P_{\delta}(\delta) =\frac{\delta^{-2}}{\cos\theta_2-\cos\theta_1}\int_{A(\delta)}^{B(\delta)}\frac{P_{\gamma}(\gamma)}{\sqrt{\gamma^2-1}}d\gamma,
\end{eqnarray}
where $P_{\gamma}(\gamma)$ is the PDF for $\gamma$. Little is known about the form of $P_{\gamma}(\gamma)$. In previous works \citep[e.g.,][]{2003ApJ...599..105L,2008ApJ...674..111C,2012ApJ...751..108A}, a power-law form with index k was usually assumed:
\begin{eqnarray}
\label{p_gamma}
P_{\gamma}(\gamma)=C\gamma^k,
\end{eqnarray}
where $C$ is a normalization constant and the function is valid for $\gamma_1 \leq \gamma \leq \gamma_2$. In this work, we also test a form similar to the relativistic Maxwell-J\"{u}ttner distribution \citep[e.g.,][]{2016ApJ...821...77K} for $P_{\gamma}(\gamma)$. In physics, the Maxwell-J\"{u}ttner distribution is the distribution of speeds of particles in a hypothetical gas of relativistic particles. We have
\begin{eqnarray}
\label{p_gamma_maxwell}
P_{\gamma}(\gamma)=\frac{\gamma\sqrt{\gamma^2-1}\exp(-\gamma/k)}{kK_2(1/k)},
\end{eqnarray}
where $k$ is a free parameter, and $K_2$ denotes the modified Bessel function of the second kind. This function is valid for $1.0<\gamma<\infty$. In practical calculation, we take a range of $1.01 \leq \gamma \leq 100$, which can ensure a good normalization.

Given the $P_{\delta}(\delta)$ and utilizing Equation (\ref{delta_p}), it is easy to determine the conditional probability distribution of $\log\mathcal{L}_{c}$ given $\log L_{c}$. In Appendix \ref{Acpd}, we give the formula of $p(\log \mathcal{L}_{c} \mid \log L_{c})$ for two cases: $q$ is a constant (Equation (\ref{beamed_condition1})), and $q$ follows the Gaussian distribution (Equation (\ref{beamed_condition2})). A Monte Carlo simulation suggests that the two cases give similar results. In the following analysis, we adopt the first case for its simplicity, and
\begin{eqnarray}
\label{beamed_condition}
p(\log \mathcal{L}_{c} \mid \log L_{c})= \frac{\ln10}{q}(\frac{\mathcal{L}_c}{L_c})^{\frac{1}{q}}P_{\delta}\left((\frac{\mathcal{L}_c}{L_c})^{\frac{1}{q}}\right).
\end{eqnarray}

\begin{table}[!t]
\tablewidth{0pt}
\renewcommand{\arraystretch}{1.3}
\caption{Parameters of the Beaming Models.}
\begin{center}
\begin{tabular}{lccc}
\hline\hline

\colhead{Parameter} & \colhead{Model $1$} & \colhead{Model $2$} & \colhead{Model $3$} \\
\hline
$q$          &  4.679$_{-0.005}^{+0.003}$    & 4.679$_{-0.017}^{+0.006}$ & 3.606$_{-0.103}^{+0.334}$  \\
$k$          &  -1.38$_{-0.16}^{+0.10}$      & 3.38$_{-0.04}^{+0.03}$    & 3.23$_{-0.85}^{+1.21}$ \\
$\gamma_1$   &  3.54$_{-0.10}^{+0.09}$       & 1.01                      & 1.01                      \\
$\gamma_2$   &  34.82$_{-3.95}^{+12.52}$     & 100                       & 100                       \\
$\theta_1$   &  40                           & 40                        & 44.78$_{-6.65}^{+6.61}$  \\
$\theta_2$   &  14                           & 14                        & 7.98$_{-0.42}^{+1.58}$   \\
\hline
\end{tabular}
\end{center}
~~~~~$\mathbf{Notes}$. Parameters without an error estimate were kept fixed during the fitting stage. The units of $\theta_1$ and $\theta_2$ are degrees.
\label{tab:beam}
\end{table}

\begin{figure*}
  \centerline{
    \includegraphics[scale=0.78,angle=0]{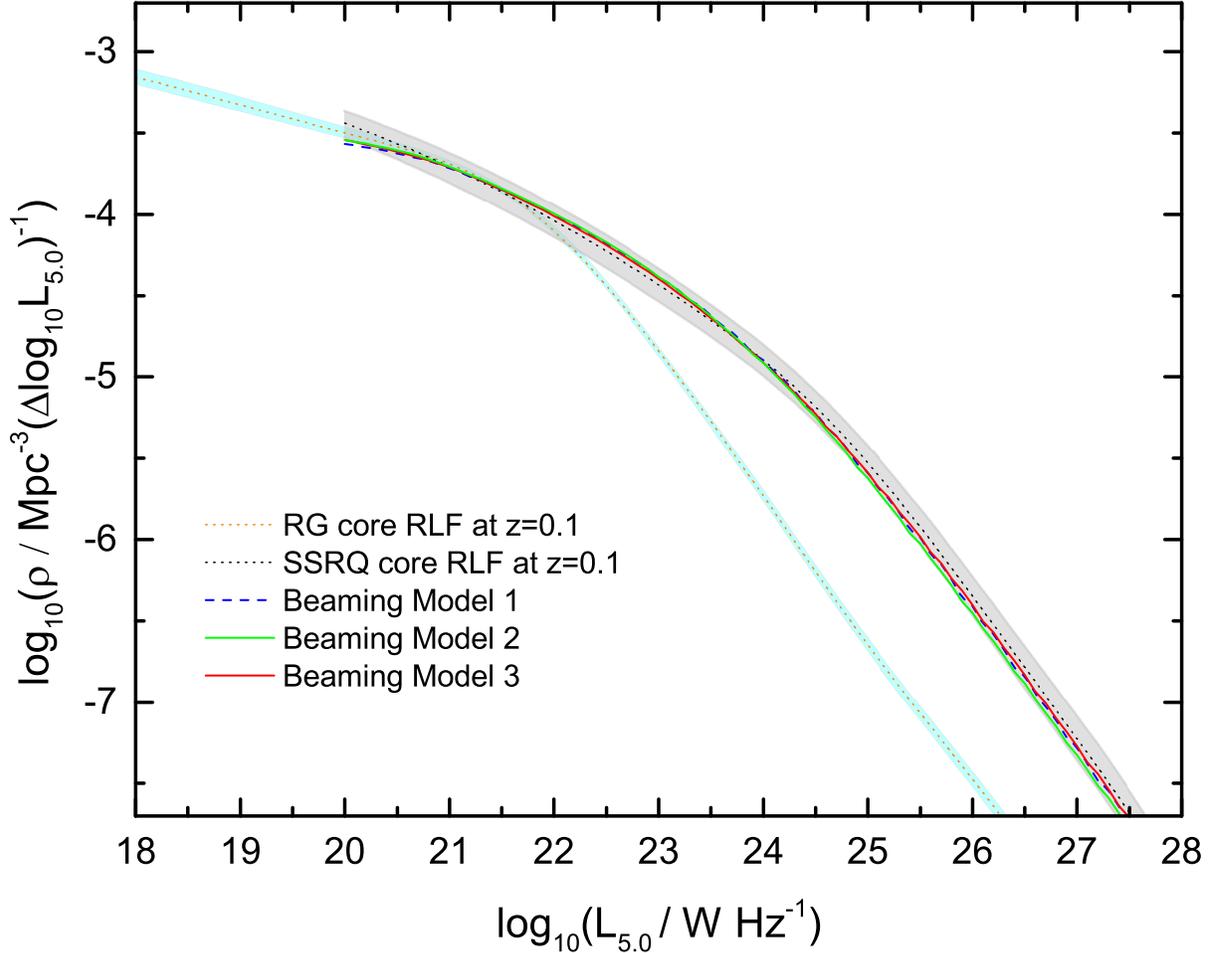}}
  \caption{\label{beaming_model} Core RLFs of RGs (orange dotted line) and SSRQs (black dotted line) at $z=0.1$ and the best-fit beaming models described in Section \ref{Intrinsic}. The light shaded areas represent the uncertainties due to the incompletness of spectral indices.}
\end{figure*}

\begin{figure}
  \centerline{
    \includegraphics[scale=0.43,angle=0]{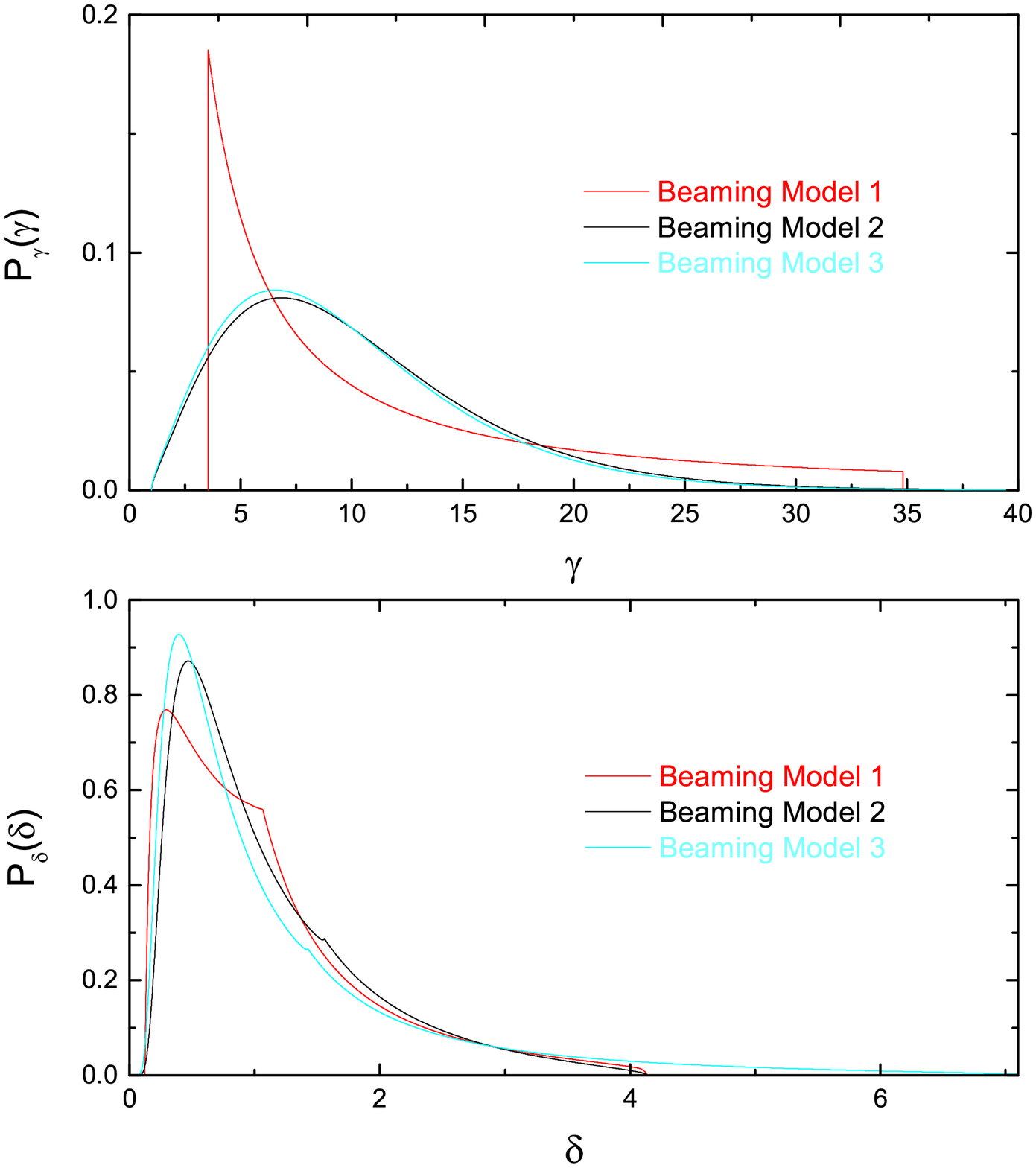}}
  \caption{\label{pdf_delta_gamma} Distributions of Lorentz factors (upper panel) and Doppler factors (lower panel) predicted by the beaming models.}
\end{figure}

Now similar to Equation (\ref{eqn:core RLF}), the Doppler beamed RG core RLF is calculated as
\begin{eqnarray}
\label{beamed_core_RLF}
\phi_c(z,\mathcal{L}_c)=\int p(\log \mathcal{L}_{c} \mid \log L_{c})\rho_{c}(z,L_{c})d\log L_c,
\end{eqnarray}
where the limits of integration are $\log L_{c,\mathrm{min}}=18$ and $\log L_{c,\mathrm{max}}=28$, roughly corresponding to the $L_c$ range for the RG sample.
By fitting Equation (\ref{beamed_core_RLF}) to the SSRQ core RLF, we can determine the parameters of the Lorentz-factor distribution, and the best-fit value of $q$. To get more information on the parameters, we use the Markov chain Monte Carlo (MCMC) sampling algorithm \citep{2002PhRvD..66j3511L}. The fitting is performed on three beaming models: (1) a power-law form for $P_{\gamma}(\gamma)$; (2) a form similar to the relativistic Maxwell-J\"{u}ttner distribution for $P_{\gamma}(\gamma)$; (3) the same form of $P_{\gamma}(\gamma)$ as model 2, but setting $\theta_1$ and $\theta_2$ as free parameters. The fit values are summarized in Table \ref{tab:beam}. The posterior probability distributions and two-dimensional (2D) confidence contours of parameters in our beaming models are given in Figure \ref{fig_mcmc}. With the 2D contours, one can inspect the degeneracies between the input parameters \citep[e.g.,][]{2016MNRAS.456.2173Y}.

Figure \ref{beaming_model} shows how the best-fit beaming models reproduce the core RLF of SSRQs. It seems that all the three models are applicable. Model 2 has fewer free parameters than Model 1. Having the same number of free parameters as Model 1, Model 3 has the advantage of constraining the range of viewing angles. It gives values of $\theta_1=44.8_{-6.7}^{+6.6}$ degrees and $\theta_2=8.0_{-0.4}^{+1.6}$ degrees. The value of $\theta_2$ is slightly smaller than that of $14^{\circ}$ given by \citet{1992ApJ...387..449P}. According to the unification scheme of AGNs, $\theta_1$ marks the division between RGs and SSRQs, and $\theta_2$ is the demarcation angle between FSRQs and SSRQs. From the relative numbers between RGs and quasars, \citet{1989ApJ...336..606B} concluded that $\theta_1=44.4^{\circ}$, very close to our result. Based on the monitoring observations with the Very Long Baseline Array (VLBA), \citet{2010AA...512A..24S} reported that 44 of 45 FSRQs in their sample have viewing angles $\leqslant 8.5^{\circ}$, and only one has a viewing angle of $14.8^{\circ}$. Their statistics are in good agreement with the results of our analysis.

In Figure \ref{pdf_delta_gamma}, we show the distributions of Lorentz factors and Doppler factors predicted by the beaming models. The power-law index of Model 1 is $k=-1.38_{-0.16}^{+0.10}$, which is in agreement with $k \thicksim -1.5$ found for the CJ-F survey \citep{1997ApJ...476..572L}. Model 1 implies an average Lorentz factor for SSRQs of $\gamma=11.68_{-0.70}^{+1.59}$. Model 2 and 3 give $\gamma=10.27_{-0.13}^{+0.10}$ and $\gamma=9.84_{-2.50}^{+3.61}$, respectively. On average, our result is close to the average Lorentz factor for Fermi-detected FSRQs, which is $\gamma=11.7_{-2.2}^{+3.3}$ given by \citet{2012ApJ...751..108A}.

\section{Discussion}
\label{discussion}

\begin{figure}
  \centerline{
    \includegraphics[scale=0.40,angle=0]{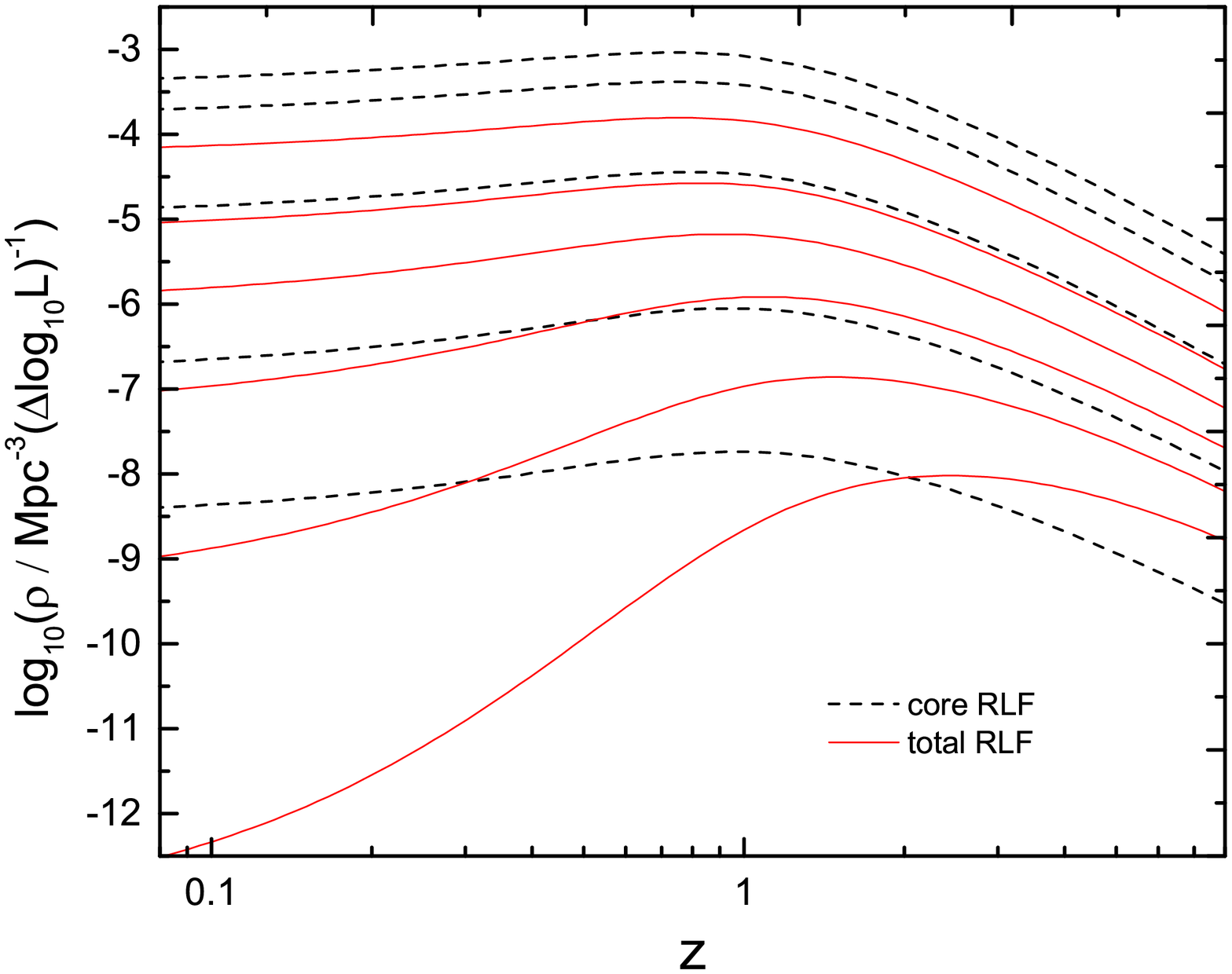}}
  \caption{\label{core_total_RLF} Comparison between core RLF for RGs and total RLF (Model A of Y17). From top to bottom, The black dashed lines show the core RLFs at $\log_{10}L_{5.0\mathrm{GHz}}$=19, 21, 23, 25 and 27, respectively, and the red solid lines represent the total RLFs at $\log_{10}L_{408\mathrm{MHz}}$=23.0, 24.5, 25.5, 26.5, 27.5 and 28.5 respectively.}
\end{figure}

\begin{figure}
  \centerline{
    \includegraphics[scale=0.35,angle=0]{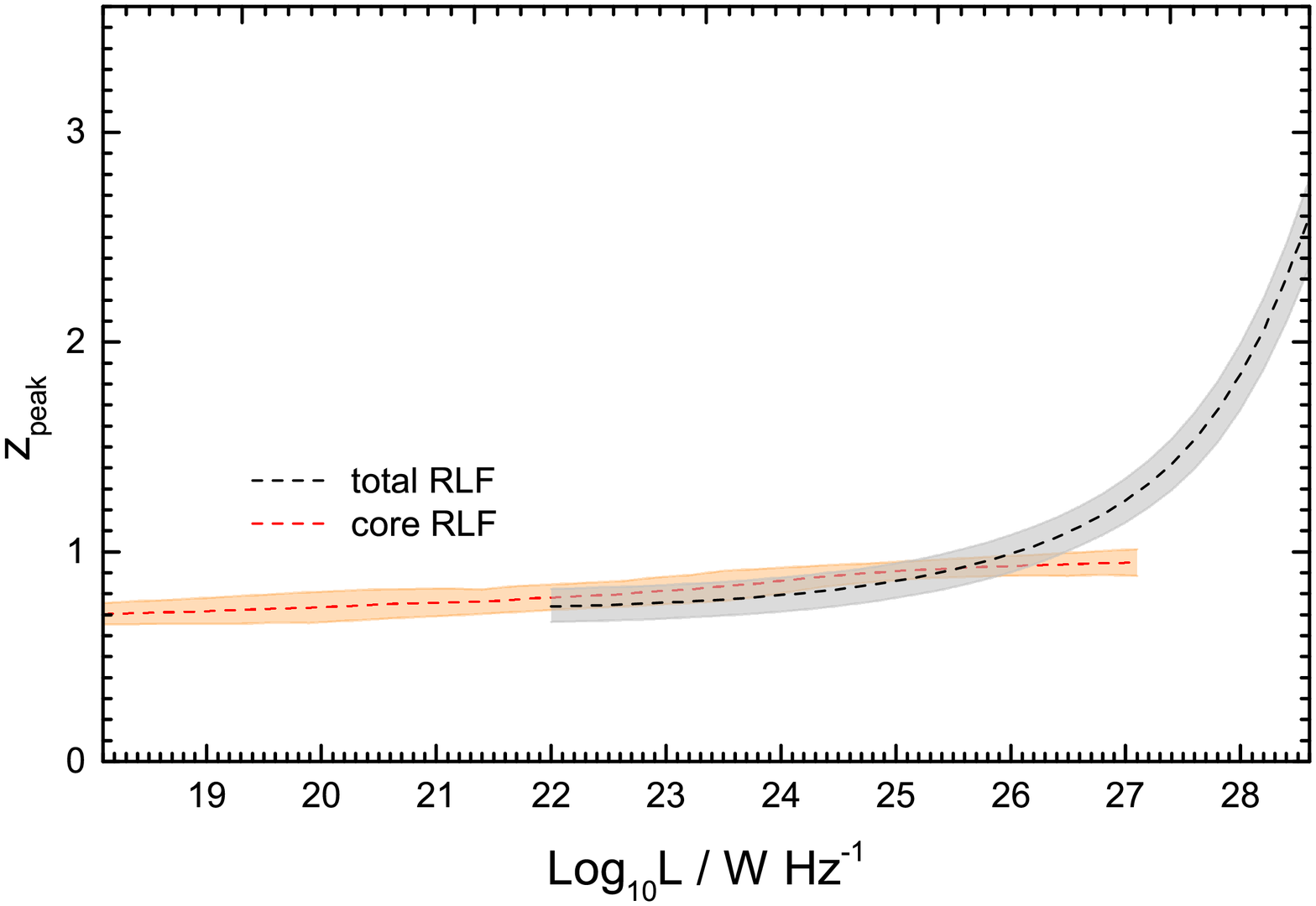}}
  \caption{\label{zpeak} Variation in the redshift of the peak space density with radio luminosity for the core RLF, compared with that for total RLF.}
\end{figure}

\begin{figure}[!h]
\centering
\includegraphics[width=1.05\columnwidth]{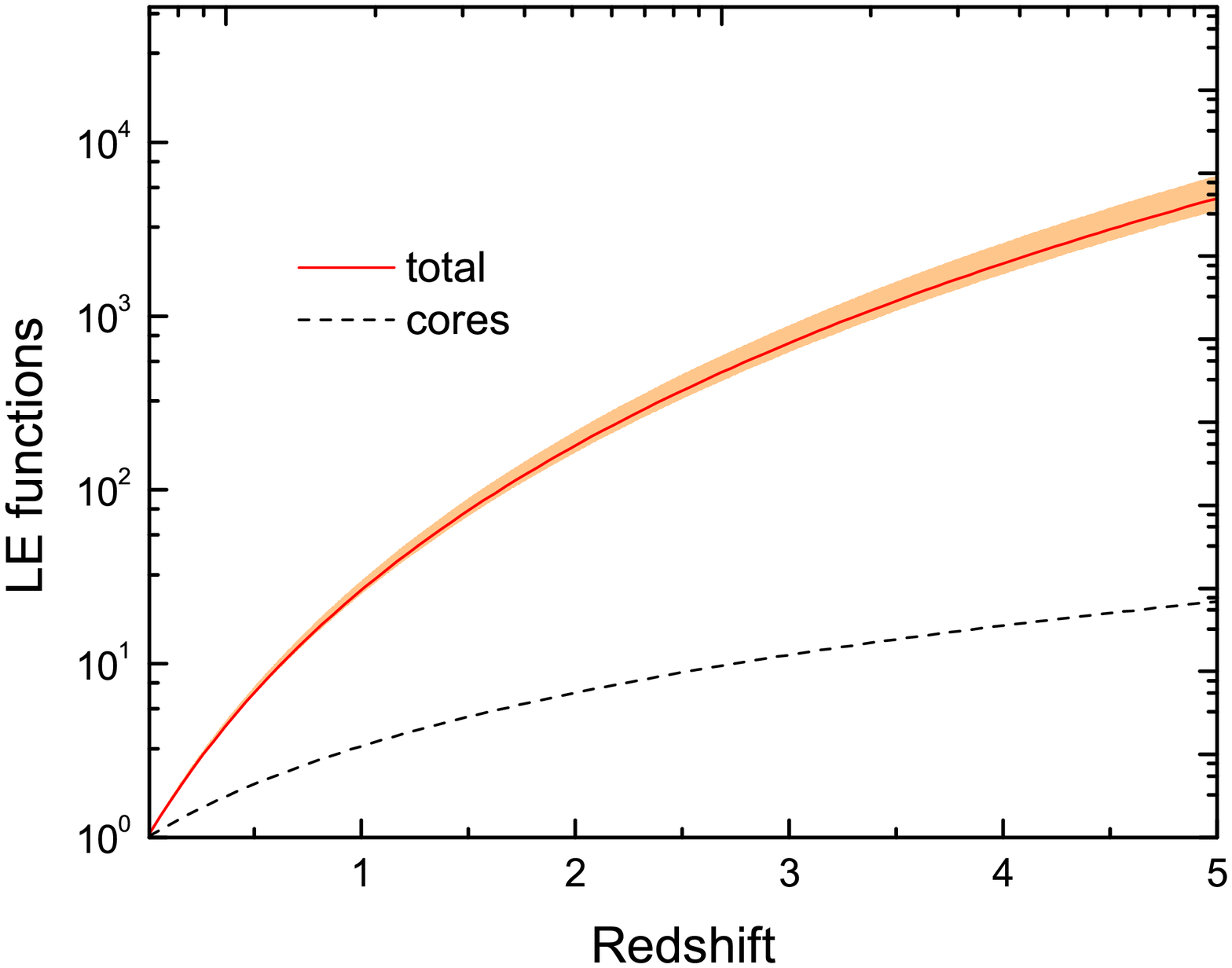}
\includegraphics[width=1.05\columnwidth]{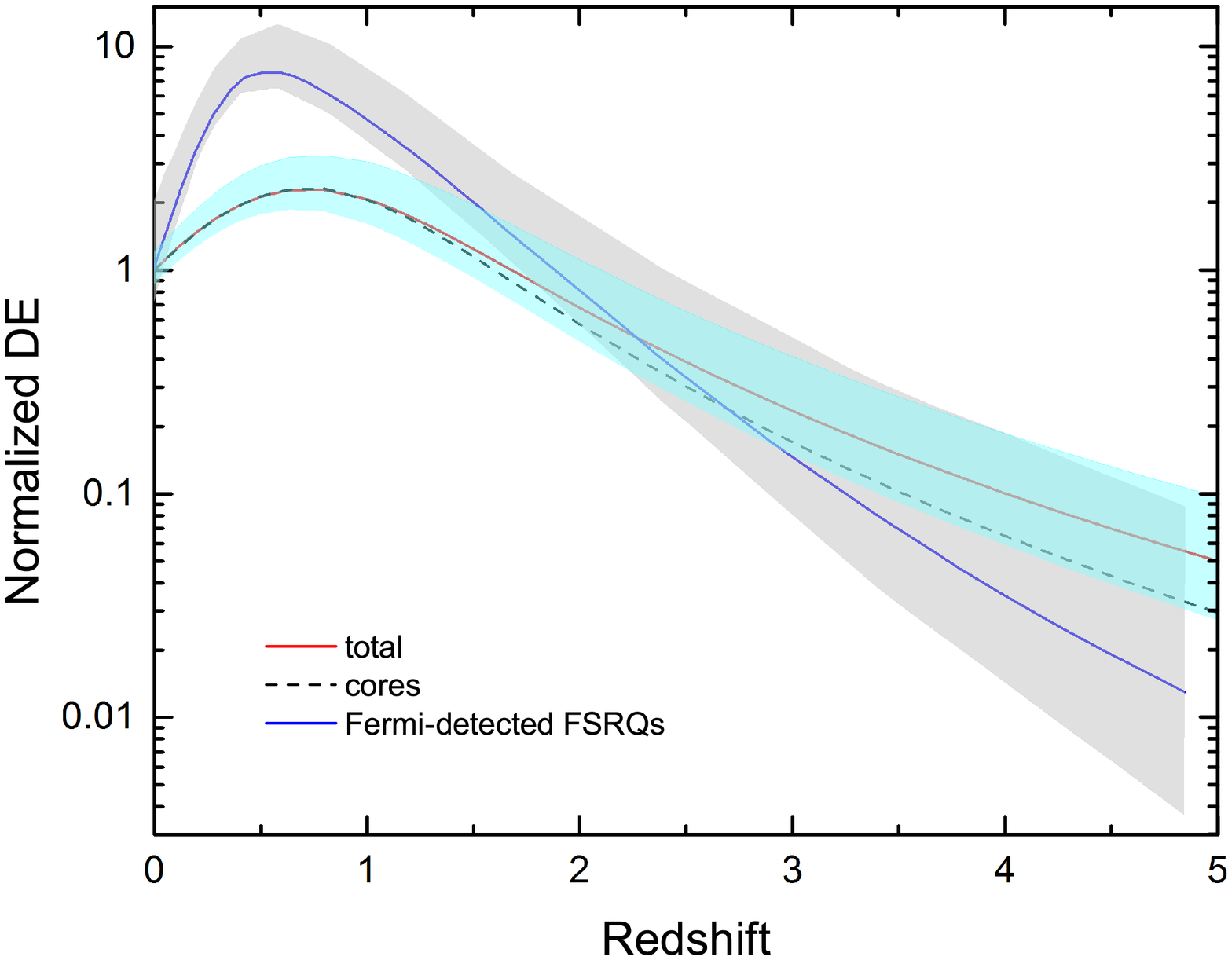}
\caption{LE (upper panel) and normalized DE (lower panel) as functions of redshift. The red solid and black dashed lines represent the total and cores, respectively. In the lower panel also plotted is the normalized DE of Fermi-detected FSRQs (blue solid line) as a function of redshift \citep[derived from the Figure 15 of][]{2012ApJ...751..108A}. The light shaded areas take into account the 1 $\sigma$ error bands.}
\label{LE_DE}
\end{figure}

\subsection {Comparing Core RLF with Total RLF}
Compared with the total RLF, the typical characteristic of the core RLF (see Figure \ref{RLFz}) is the negative evolution occurring at a redshift of $z\gtrsim0.8$. In Figure \ref{core_total_RLF}, we plot the core RLF for RGs and the total RLF (also see the ``Model A'' panel of Figure 3 in Y17) together. Note that no matter for low- or high-luminosity cores, the variation of space density with redshift behaves very similarly, implying a very weak luminosity-dependent evolution. As for the total RLF, however, both the amount of space density changing from redshift zero to the maximum space density, and the peak redshift are strong functions of radio luminosity. Figure \ref{zpeak} shows the variation in the redshift of the peak space density with radio luminosity for the core RLF, compared with that for total RLF. Note for the core RLF, the peak redshift increases very slightly with radio core luminosity, while for the total RLF, the increase is dramatic.

The parametric core RLF in Section \ref{parametric_rhoc} allows us to determine the DE and LE for radio cores. They are given by Equations (\ref{ez1}) and (\ref{ez2}). In the upper panel of Figure \ref{LE_DE}, we plot the LE function of radio cores compared with that of the total source. Both the cores and total source show a positive LE, but the LE of the cores is less dramatic. The positive LE suggests that both the radio cores and lobes at higher redshift are systematically brighter than those of today. A possible explanation is that both the average density of the universe and the gas fraction are higher \citep{2014MNRAS.445..955B} at higher redshifts, so that the radio lobes of AGNs remain more confined and adiabatic expansion losses are lower, leading to higher synchrotron luminosities \citep[e.g.,][]{1996MNRAS.283L..45B}. On the other hand, the positive LE for cores is milder than that for lobes, implying that the denser environment at high-redshift has relatively less impact on the core luminosity. Less interaction with external environment often means less shocks, less energy dissipation, and less radio emission \citep{2004NewAR..48.1157F}.

The DE function of the cores can not be compared directly with that of the total source. We define the normalized DE function as:
\begin{eqnarray}
\label{nde}
\varrho(z)=\int_{L_{min}}^{L_{max}}\rho(z,L)dL/\int_{L_{min}}^{L_{max}}\rho(z=0,L)dL.
\end{eqnarray}
The normalized DE functions of radio core and total source are shown
in the lower panel of Figure \ref{LE_DE}. The two functions
  are in good agreement within the uncertainty range,
indicating that the core and lobes co-evolve with
  redshift. It is possible that they are not completely
  consistent, e.g., episodic AGN activity could cause
  deviations. This would allow the presence of RGs with a
  ``switched-off¡± core \citep[e.g.,][]{2009AA...506L..33M}, or having
  dying radio lobes from an earlier active epoch along with newly
  restarting jets and cores \citep[e.g.,][]{2011AA...526A.148M}. But
such sources don't appear to dominate our sample.

\citet{1995AA...298..375F} argued that the difference between radio
loud and radio weak is established already on the parsec scale. We find that the DE
function of radio cores peaks at $z \thicksim 0.8$ and then rapidly
decreases, indicating that core-bright radio-loud AGNs at high
redshift are less numerous. The redshift at which radio cores peak
is lower than the redshift of BH growth. The reason for this is
  not entirely clear but it is presumably related to
  redshift-dependent accretion efficiency and jet triggering. For
  example, simulations of AGN evolution by \citet{2014MNRAS.442.2304H}
  have revealed that the number of BHs accreting close to the Eddington rate decrease with increasing redshift. This implies that the dominant mechanism of AGN fuelling changes with cosmic time from cold gas accretion via major mergers to radiatively inefficient accretion directly from hot gas haloes \citep{2015AA...581A..96R}.

In the lower panel of Figure \ref{LE_DE}, we also plot the normalized DE of Fermi-detected FSRQs as a function of redshift \citep[adopted from the Figure 15 of][]{2012ApJ...751..108A}. The general trend of their result is consistent with our determination. Nevertheless, it seems that both the decline in the space density after the redshift peak and the increase in space density leading up to the redshift peak are more dramatic than that of our result. It is also noticed that their peak redshift is $z\thicksim0.6$, being smaller than our determination of $z\thicksim0.8$. We speculate that the above difference is caused because the Fermi-detected FSRQ sample bias to those extreme FSRQs with, on average, faster apparent jet speeds and smaller viewing angles \citep[e.g.,][]{2009AJ....138.1874L,2010AA...512A..24S}. It represents an extreme sub-sample of FSRQs.

\begin{figure}
  \centerline{
    \includegraphics[scale=0.40,angle=0]{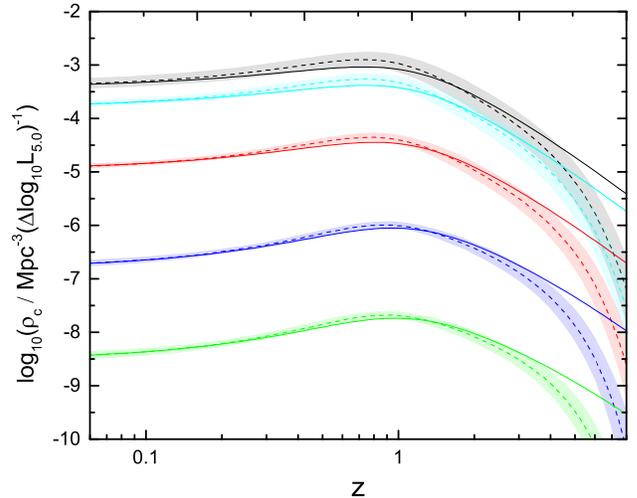}}
  \caption{\label{new_total_RLF} Core RLF for RGs determined based on a different total RLF model (the Model C of Y17). The black, cyan, red, blue and green dashed lines show the core RLFs at $\log_{10}L_{5.0\mathrm{GHz}}$=19, 21, 23, 25 and 27, respectively. The light shaded areas take into account the uncertainties due to the incompletness of spectral indices, as well as the 1 $\sigma$ error propagated from the Model C total RLF by Y17. The solid curves represent the core RLF for RGs determined in Section \ref{Semi_parametric_rhoc}.}
\end{figure}

\subsection{Other input models of total RLF}

Our key equation for determining the core RLF is given by Equation (\ref{eqn:core RLF}). Given $p(\log L_{c} \mid \log L_{t})$, the calculation of core RLF depends on the model adopted for the total RLF. In order to rule out the possibility that a different total RLF model may significantly change the main result, we need to perform a comparison test. In the test, we adopt the Model C of Y17 as the new total RLF, for which $\rho_{t}(z,L_{t})$ and $e_1(z)$ are also given by Equation (\ref{rhototal}) and (\ref{ez1}), while $e_2(z)$ is given by
\begin{eqnarray}
\label{otherez2}
e_2(z)=10^{k_1z^2+k_2z}.
\end{eqnarray}
Model C permits the possibility of negative LE at high redshift, and it was comparable to Model A in fitting the data of Y17.
Figure \ref{new_total_RLF} compares the core RLFs derived for the two total RLF models. The black, cyan, red, blue and green dashed lines show the core RLFs at $\log_{10}L_{5.0\mathrm{GHz}}$=19, 21, 23, 25 and 27 respectively for Model C. The solid curves represent the core RLF for RGs determined in Section \ref{Semi_parametric_rhoc}. The core RLFs are not significantly different at $z\lesssim3$. Their only difference lies in the steepness of the high-redshift decline of $\rho_c$. Due to lack of high-redshift samples, the total RLFs in Y17 cannot conclude whether the high-redshift decline of $\rho_t$ is sharp or shallow. The core RLFs here inherit such uncertainty.

\subsection{Luminosity-dependent evolution}
In the past decades, it has became well established that the evolution of the luminosity functions (LFs) of AGNs is luminosity-dependent \citep[e.g.,][]{2001MNRAS.328..882W,2003ApJ...598..886U,2005AA...441..417H,2007ApJ...654..731H,2009MNRAS.399.1755C,2010MNRAS.401.2531A,2011MNRAS.416.1900R,2012ApJ...751..108A,2013MNRAS.431..997Z,2014MNRAS.439.2736D}.
Physically, this was usually interpreted as a sign of cosmic downsizing, where the most massive black holes form at earlier epochs than their less massive counterparts \citep{2015AA...581A..96R}.
To describe the Luminosity-dependent evolution of optical, X-ray and $\gamma$-ray LFs, the luminosity-dependent density evolution (LDDE) model was developed and became popular. But the LDDE model is unable to model the steep-spectrum RLF (Y17). We thus develop a mixture evolution scenario \citep[][Y17]{2016ApJ...820...65Y} which suggests that the evolution of RLF is due to a combination of DE and LE. In essence, the DE determines when the density curve will peak and when it will decline, while the LE can shift the location of peaks according to different luminosities, such that a luminosity-dependent evolution is a natural result. The mixture evolution scenario is especially suitable for interpretation of the difference between core and total RLFs: although the cores and lobes experience synchronous DE, the cores have significantly weaker LE than lobes.

\subsection {Comparison with Previous Studies}
\begin{figure}
  \centerline{
    \includegraphics[scale=0.42,angle=0]{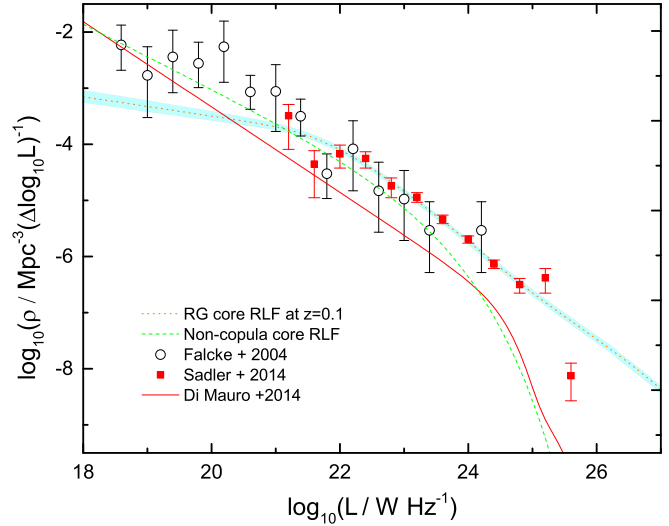}}
  \caption{\label{previous_RLF} Comparison of our core RLFs with previous results. The orange dotted line shows our RG core RLF at $z=0.1$, with the cyan band taking into account the combined uncertainty due to the incompletness of spectral indices, as well as the 1 $\sigma$ error propagated from the total RLF by Y17. The green dashed line shows the core RLF derived by non-copula method. The black open circles with error bars represent the 15 GHz core RLF of nearby galaxies measured by \citet{2004NewAR..48.1157F}. The local RLF of RGs at 20 GHz measured by \citet{2014MNRAS.438..796S} is shown as red solid squares. The red solid line shows the core RLF derived by \citet{2014ApJ...780..161D}. A flat spectrum for the core ($\alpha_c=0$) is assumed, ensuring that core RLFs at different frequencies can be compared directly.}
\end{figure}

In the decades since the discovery of radio AGN, studies on the core RLF have been few. An early determination of the core RLF was given by \citet{2004NewAR..48.1157F}. Based on the 150 mas-scale radio nuclei in the Palomar sample, they derived the 15 GHz core RLF of nearby galaxies (mainly consist of low-luminosity AGNs). This result is shown as black open circles with error bars in Figure \ref{previous_RLF}. Note that \citet{2004NewAR..48.1157F}'s errors are large and within those errors his core RLF is in reasonable agreement with our result. Nevertheless, at the faint end ($\log_{10}L_{5.0\mathrm{GHz}}<21$) his core RLF appears to be higher than our RG core RLF. This is because at the faint end our core RLF may not sufficiently consider the contribution of low-luminosity AGNs.

Based on a combined sample of steep-spectrum radio AGNs, \citet{2012ApJ...744...84Y} investigated the core RLF using the binned $1/V_{max}$ method. However, that core sample was not strictly a flux limited complete sample, and the minimum core flux density of the sample was used as the flux limit to estimate $1/V^{i}_{max}$. This would significantly underestimate the core RLF \citep{2013Ap&SS.345..305Y}. Thus the result in that work can only be regarded as a rough estimation. \citet{2012ApJ...744...84Y} concluded that the comoving number density of radio cores displays a persistent decline with redshift, implying a strong negative evolution. Now it seems that this conclusion partly reflects the truth. The result based on the more rigorous method in this work indicates that the negative evolution of cores occurs at a redshift of $z\gtrsim0.8$.

Using a sample of 202 radio sources from the Australia Telescope 20 GHz (AT20G) survey identified with galaxies from the 6dF Galaxy Survey (6dFGS), \citet{2014MNRAS.438..796S} made the first measurement of the local RLF of RGs at 20 GHz. Since the radio emission from active galaxies at 20 GHz arises mainly from the galaxy core, rather than from extended radio lobes \citep[e.g.,][]{2006MNRAS.371..898S}, the measurement of \citet{2014MNRAS.438..796S} can be treated as the local core RLF. In Figure \ref{previous_RLF}, their result is shown as red solid squares, and is in good agreement with our core RLF.

Another study involving the core RLF was performed by \citet[][hereafter D14]{2014ApJ...780..161D}. They obtained the core RLF from the total RLF of \citet{2001MNRAS.322..536W} by a simple transformation:
\begin{eqnarray}
\label{dimauro}
\rho_c(z,L_c)=\rho_{t}(z,L_{t}(L_{c})) \frac{d\log L_t}{d\log L_c},
\end{eqnarray}
where $L_{t}(L_{c})$ and $d\log L_t/d\log L_c$ derive from the total-core correlation, i.e.,
\begin{eqnarray}
\label{dimauro_Lc_Lt}
\log L^{5GHz}_c=(4.2\pm2.1)+\log L^{1.4GHz}_t(0.77\pm0.08).
\end{eqnarray}
The premise of using Equation (\ref{dimauro}) is that $L_c$ is a function of $L_t$. But obviously, there is no definite functional relationship between $L_c$ and $L_t$. The only rigorous concept describing the correlation between $L_c$ and $L_t$ is conditional probability, which can be well measured via the copula method, while the linear fit like Equation (\ref{dimauro_Lc_Lt}) is only a rough sketch. Thus the estimation obtained with Equations (\ref{dimauro}) and (\ref{dimauro_Lc_Lt}) may distort the true core RLF. In Figure \ref{previous_RLF}, we show the core RLF derived by D14 as red solid line.

\subsection {Copula versus non-copula method}
Both the copula approach here and D14's simpler approach are
  indirect techniques of estimating the core RLF. The precision
  significantly depends on how accurately the $L_C-L_T$ correlation is
  measured. Unlike our copula method, D14's approach
  does not incorporate the intrinsic dispersion in the
  $L_C-L_T$ correlation. To further compare the core RLF derived by
  copula with that using a non-copula approach, we derive
the core RLF by applying D14's transformation approach to our
  total RLF (Y17 model A). This is shown as the green
  dashed line in Figure \ref{previous_RLF}. Note that D14's core RLF
  and our non-copula core RLF agree, but they are
  significantly different from the core RLF derived using the
  copula method. They are steeper at both faint
  ($\log_{10}L_{5.0\mathrm{GHz}}<21$) and bright
  ($\log_{10}L_{5.0\mathrm{GHz}}>24$) luminosities. In general, they
  are inferior to the copula-based result in fitting the observed
  data, particularly those data obtained more recently.

\section[]{Summary and conclusions}
\label{sum}

The main results of this work are as follows.

\begin{enumerate}
  \item We verified, through a partial correlation analysis, that the correlation between the core and total radio luminosities of radio AGNs is significant. We then explored the correlation via a powerful statistical tool called ``Copula". For both RGs and SSRQs, we find that the number 13 Archimedean copula of \citet{Nelson} can well describe the $L_{c}-L_{t}$ correlation. Our results find the copula is tail independent, implying that when the cores reach extreme luminosities, the probability that lobes also show extreme luminosities tends to zero.

  \item The conditional probability distribution $p(\log L_{c} \mid
    \log L_{t})$ is obtained based on the copula-described
    $L_{c}-L_{t}$ relation. We then derive the core radio
      luminosity functions as a
    convolution of $p(\log L_{c} \mid \log L_{t})$ and the total RLF
    which was determined by \citet{2017ApJ...846...78Y}. The core RLFs
    are derived separately for RGs and SSRQs according to their own
    copula description. Our results are in reasonable agreement with
      studies that have  used radio emission at high frequency as a
      measure of the core emission.

  \item We argue that for a specific population of AGNs (e.g., SSRQs), the jet angles should be (randomly) distributed within $\theta_2 \leq \theta \leq \theta_1$ but not necessarily $0^{\circ} \leq \theta \leq 90^{\circ}$. Thus the formula calculating the PDF $P_{\delta}(\delta)$ for $\delta$ by \citet{2003ApJ...599..105L} should be modified to apply to more general conditions. In this work we give the generalized formula for deriving $P_{\delta}(\delta)$.

  \item By assuming that the RG core RLF is the intrinsic core RLF, we find the SSRQ core RLF can be reproduced by imposing a Doppler beaming effect on the RG core RLF. Consistent with previous studies, we find that the distribution of Lorentz factor can be described by a power-law form, and a form similar to the relativistic Maxwell-J\"{u}ttner distribution is also applicable. Our preferred beaming model suggests that SSRQs have an average Lorentz factor of $\gamma=9.84_{-2.50}^{+3.61}$, and that most are seen within $8^{\circ} \lesssim \theta \lesssim 45^{\circ}$ of the jet axis.

  \item We find that while the density evolution of the core and total
    RLFs match within uncertainties, there is a significant
    difference in their luminosty evolution. The core RLF
    presents a very weak luminosity-dependent evolution, with the
    number density peaking around $z\thicksim 0.8$ for all
    luminosities. The redshift at which core RLF peaks is lower than that of the peak of BH growth. The reason for this is not entirely clear but it is presumably related to redshift-dependent accretion efficiency and jet triggering.

\end{enumerate}

\acknowledgments

We are grateful to the referee for useful comments that improved this paper. We acknowledge the financial support from the National Natural Science Foundation of China 11603066, U1738124, 11573060 and 11661161010. We would like to thank Xiaolin Yang, Ming Zhou, Guobao Zhang and Dahai Yan for useful discussions. JM is supported by the Hundred Talent Program, the Major Program of the Chinese Academy of Sciences (KJZD-EW-M06), the National Natural Science Foundation of China 11673062, and the Overseas Talent Program of Yunnan Province. BZ is supported by the National Thousand Young Talents program of China. The authors gratefully acknowledge the computing time granted by the Yunnan Observatories, and provided on the facilities at the Yunnan Observatories Super-computing Platform. This research has made use of the NASA/IPAC Extragalactic Database (NED), which is operated by the Jet Propulsion Laboratory, California Institute of Technology, under contract with the National Aeronautics and Space Administration.

\clearpage

\appendix

\section{The sample of 503 sources}\label{details}

\begin{deluxetable}{lllrcrrcc}
\tablecolumns{9}
\tabletypesize{\footnotesize}
\tablewidth{0.99\textwidth}
\tablecaption{Summary of Sample$^{a}$\label{tab:sample}}
\tablehead{
\colhead{IAU}   &
\colhead{Other}  &
\colhead{$z$}  &
\colhead{$S_{t0.408}$}  &
\colhead{$\alpha_{t}$} &
\colhead{$S_{core5.0}$} &
\colhead{$\alpha_{c}$}      &
\colhead{Classification}      &
\colhead{References} \\
\colhead{}   &
\colhead{Name}   &
\colhead{}   &
\colhead{Jy}   &
\colhead{}      &
\colhead{mJy}  &
\colhead{} &
\colhead{} &
\colhead{}
}

\startdata
0101$-$649  &                     & 0.1630   &    1.15        &   0.55  &      179.2   & -0.22   & G   & 1         \\
0736$+$017  &                     & 0.1910   &   2.840        &   0.21  &     1780     &         & Q   & 2         \\
2315$-$425  &  PMN J2317$-$4213   & 0.0560   &    0.97$^i$    &   0.80  &  $<$  20.6   &  0.49   & G   & 1,37      \\
2316$-$423  &                     & 0.0545   &    1.67        &   0.05  &      139.9   &  0.1    & G   & 1         \\
0123$-$016  &                     & 0.0180   &   16.40        &   0.93  &      100     & -0.3    & G   & 2         \\
0005$-$199  &                     & 0.1223   &    2.08        &   0.70  &       14     & -0.54   & G   & 3,4,5     \\
0222$+$36   &                     & 0.0327   &    0.37        &   0.21  &      140     & -0.47   & G   & 6,7,8     \\
1144$+$352  &  B2 1144$+$35B      & 0.0631   &    0.33        &  -0.53  &      243     &         & G   & 7,18,25   \\
2308$+$098  &  4C09.72            & 0.432    &    1.99        &   0.74  &      102     &         & Q   & 21,23,24   \\
\enddata
\tablenotetext{a}{Table \ref{tab:sample} is available in its entirety in machine-readable forms in the online journal.
A portion is shown here for guidance regarding its form and content.}
\tablecomments{
  Column (1). Source name in IAU designation (B1950).
  Column (2). Other name if available.
  Column (3). Spectroscopic redshift.
  Column (4). Total flux density at 408 MHz in Jy. Those with a flag ``i" mean that their $S_{t0.408}$ are interpolated from near frequencies.
  Column (5). Spectral index near 408 MHz, defined by $S \varpropto \nu^{-\alpha}$).
  Column (6). Core flux density at 5 GHz in mJy.
  Column (7). Core spectral index near 5 GHz.
  Column (8). Classification: G=radio galaxy; Q=quasar.
  Column (9). References:
  (1) \citet{1994MNRAS.268..602J}; (2) \citet{1993MNRAS.263.1023M}; (3) \citet{1989MNRAS.236..737E}; (4) \citet{1994MNRAS.269..928S}; (5) \citet{2000AA...353..507G}; (6) \citet{2003MNRAS.338..176H}; (7) \citet{2009AA...505..509L}; (8) \citet{2005AA...441...89G}; (9) \citet{1991AA...245..371B}; (10) \citet{1984AA...139...55F}; (11) \citet{1987AAS...69...57F}; (12) \citet{1978AAS...34..341F}; (13) \citet{2002AA...383..104C}; (14) \citet{1995AA...300..643C}; (15) \citet{1997AAS..126..335M}; (16) \citet{2004AA...425..825K}; (17) \citet{1988AA...199...73G}; (18) \citet{2007AA...474..409G}; (19) \citet{1999MNRAS.310...30C}; (20) \citet{1999ApJS..124..285R}; (21) \citet{1990PKS...C......0W}; (22) \citet{1981MNRAS.194..693L}; (23) \citet{1991Obs...111...72L}; (24) \citet{1998AAS..132...31N}; (25) \citet{1970AAS....1..281C}; (26) \citet{1996AJ....111.1945D}; (27) \citet{1985AAS...59..255F}; (28) \citet{1989MNRAS.238.1055R}; (29) \citet{1969ApJ...157....1K}; (30) \citet{2010MNRAS.401...67S}; (31) \citet{1998AJ....115.1693C}; (32) \citet{1990MNRAS.246..256H}; (33) \citet{1993MNRAS.264..721L}; (34) \citet{1989AJ.....97...36M}; (35) \citet{1978AA....67...47E}; (36)\citet{1992MNRAS.257..353M}; (37)\citet{1994ApJS...91..111W}; (38)\citet{1992ApJS...79..331W}
}
\end{deluxetable}

\section{Partial correlation analysis}\label{PCAs}
In statistics, partial correlation measures the degree of association between two random variables, after eliminating the effect of all other random variables.
Suppose there are three random variables $x_i$, $x_j$ and $x_k$, the correlation coefficient between two of them, say $x_i$ and $x_j$, is denoted by $r_{ij}$. The partial correlation of $x_i$ and $x_j$ given $x_k$ is \citep{1979ats..book.....K}
\begin{eqnarray}
\label{r_ij_k}
r_{ij|k}=\frac{r_{ij}-r_{ik}r_{jk}}{\sqrt{1-r_{ik}^2}\sqrt{1-r_{jk}^2}},
\end{eqnarray}
The correlation coefficients $r_{ij}$, $r_{ik}$ and $r_{jk}$ can be calculated based on Pearson's, Kendall's, or Spearman's correlation methods.
In this work we use the Spearman rank-order correlation coefficient, which is given by Equation (A1) of \citet{2011ApJ...733...66I}.
According to \citet{Kim2015}, the statistics $t_{ij|k}$ of the partial correlation is calculated by
\begin{eqnarray}
\label{t_ij_k}
t_{ij|k}=r_{ij|k}\sqrt{\frac{N-2-g}{1-r_{ij|k}^2}},
\end{eqnarray}
where $N$ is the sample size and $g$ is the total number of given variables (here $g$=1). The probability of the null hypothesis that $x_i$ and $x_j$ are uncorrelated, i.e. the \emph{p}-value is given by
\begin{eqnarray}
\label{p_value}
p_{ij|k}=2\Phi_t(-|t_{ij|k}|,N-2-g),
\end{eqnarray}
where $\Phi_t(\cdot)$ is the cumulative density function of a Student's \emph{t} distribution with the degree of freedom $N-2-g$ \citep[see][for details]{Kim2015}.

\section{Doppler factor distributions}\label{DFSs}

We determine the PDF $P_{\delta}(\delta)$ that describes the expected distributions of Doppler factors for a randomly oriented, two-sided jet population. Suppose the PDF of Lorentz factors is $P_{\gamma}(\gamma)$ which is valid for $\gamma_1 \leq \gamma \leq \gamma_2$. As mentioned in section \ref{Intrinsic}, SSRQs have their radio axes within $\theta_1 \gtrsim \theta \gtrsim \theta_2$, and $\theta_1=40^{\circ}$, $\theta_2=14^{\circ}$. Thus the viewing angles are distributed according to
\begin{eqnarray}
\label{viewing_angles}
P_{\theta}(\theta)=\frac{\sin\theta}{\cos\theta_2-\cos\theta_1},
\end{eqnarray}
We define
\begin{eqnarray}
\label{f_delta}
f_{\delta}(\gamma,\theta)=\left(\gamma-\sqrt{\gamma^2-1}\cos\theta\right)^{-1},
\end{eqnarray}
and
\begin{eqnarray}
\label{g_delta}
g_{\pm}(\delta,\theta)=\frac{1\pm \cos\theta\sqrt{1-\delta^2\sin^2\theta}}{\delta\sin^2\theta}.
\end{eqnarray}
Given $\theta_2 \leq \theta \leq \theta_1$ and $\gamma_1 \leq \gamma \leq \gamma_2$, the possible Doppler factors range from
\begin{eqnarray}
\label{delta_min}
\delta_{min}=f_{\delta}(\gamma_2,\theta_1)
\end{eqnarray}
to
\begin{eqnarray}
\label{delta_max}
\delta_{max} =
\begin{cases}
  \displaystyle f_{\delta}(\gamma_1,\theta_2),  &  \frac{1}{\sin\theta_2} < \gamma_1 \\
                f_{\delta}(\gamma_2,\theta_2),   &  \frac{1}{\sin\theta_2} > \gamma_2  \\
                \frac{1}{\sin\theta_2}, &     \gamma_1 \leq \frac{1}{\sin\theta_2} \leq \gamma_2

\end{cases}
\end{eqnarray}
According to the theory of probability transformation for several variables \citep[e.g.,][]{2003ApJ...599..105L}, the PDF for $\delta$ is given by
\begin{eqnarray}
\label{PDF_delta}
P_{\delta}(\delta) =
\begin{cases}
  \displaystyle \frac{\delta^{-2}}{\cos\theta_2-\cos\theta_1}\int_{A(\delta)}^{B(\delta)}\frac{P_{\gamma}(\gamma)}{\sqrt{\gamma^2-1}}d\gamma,  &  \delta_{min} \leq \delta \leq \delta_{max},  \\
                0, & \mathrm{elsewhere},
\end{cases}
\end{eqnarray}
where the upper limit of integral is
\begin{eqnarray}
\label{B_delta}
B(\delta)=\min\left[\gamma_2,g_{+}(\delta,\theta_2)\right].
\end{eqnarray}

\begin{figure*}
  \centerline{
    \includegraphics[scale=0.6,angle=0]{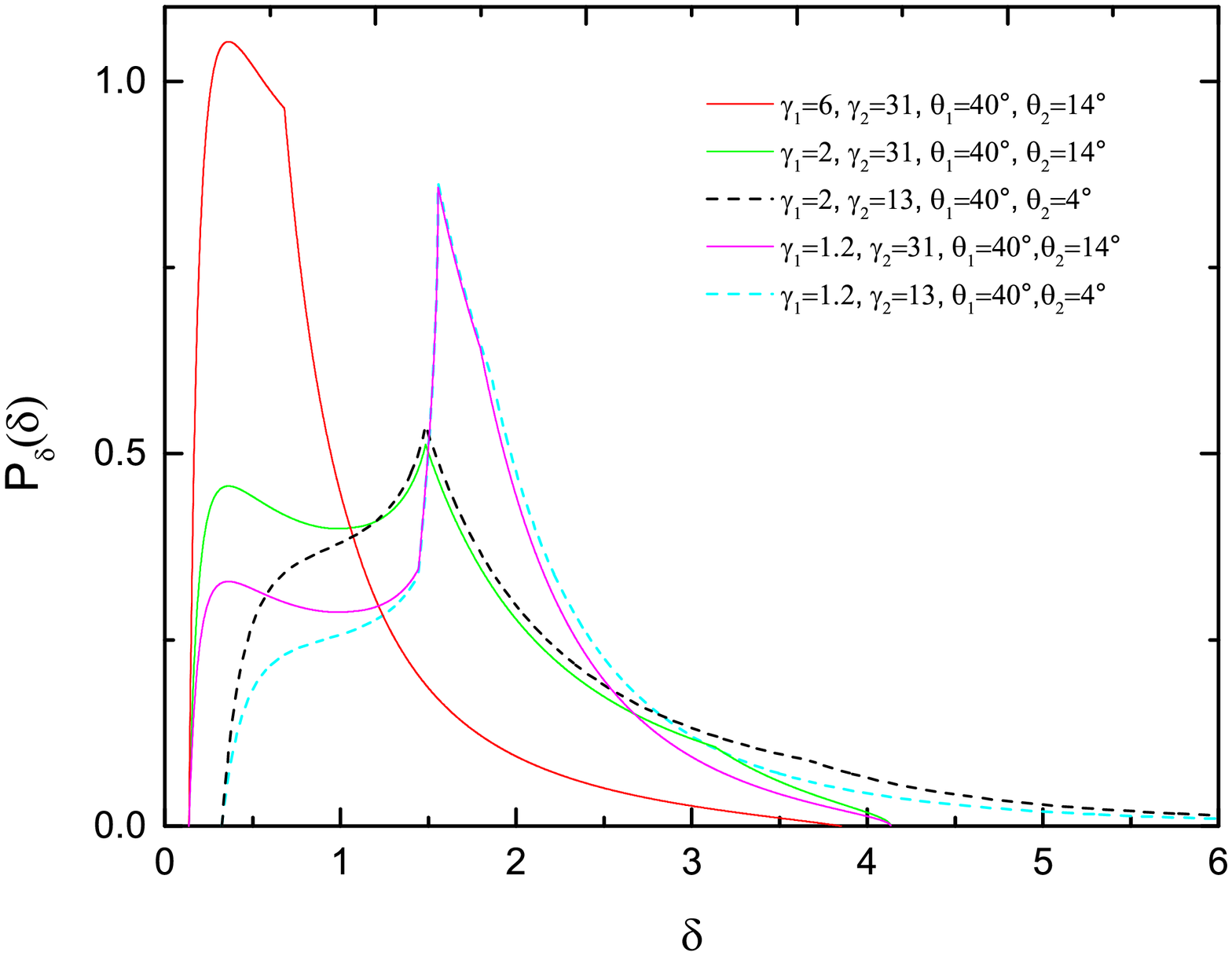}}
  \caption{\label{fA1} PDF of $\delta$ derived from Equation (\ref{PDF_delta}) with $P_{\gamma}(\gamma)\propto \gamma^{-1.5}$. The red, green, black, magenta and cyan curves show the five cases discussed from Equation (\ref{case1}) to (\ref{case5}), respectively.}
\end{figure*}

The lower limit of integral is a bit more complex than that discussed by \citet[][see their Equation A6]{2003ApJ...599..105L}. It depends on the relationship between $\gamma_1$, $\gamma_2$, $\theta_1$ and $\theta_2$.

\begin{enumerate}
  \item If $\frac{1}{\sin\theta_1}<\frac{1}{\sin\theta_2}<\gamma_1<\gamma_2$, then

\begin{eqnarray}
\label{case1}
A(\delta) =
\begin{cases}
  \displaystyle g_{+}(\delta,\theta_1),  &  \delta_{min} \leq \delta < f_{\delta}(\gamma_1,\theta_1) \\
                \gamma_1, & f_{\delta}(\gamma_1,\theta_1) \leq \delta \leq f_{\delta}(\gamma_1,\theta_2)
\end{cases}
\end{eqnarray}

  \item Else if $\frac{1}{\sin\theta_1}<\gamma_1<\frac{1}{\sin\theta_2}<\gamma_2$, then
\begin{eqnarray}
\label{case2}
A(\delta) =
\begin{cases}
  \displaystyle g_{+}(\delta,\theta_1),  &  \delta_{min} \leq \delta < f_{\delta}(\gamma_1,\theta_1) \\
                \gamma_1, & f_{\delta}(\gamma_1,\theta_1) \leq \delta < f_{\delta}(\gamma_1,\theta_2) \\
                g_{-}(\delta,\theta_2), &  f_{\delta}(\gamma_1,\theta_2) \leq \delta \leq \frac{1}{\sin\theta_2}
\end{cases}
\end{eqnarray}

\item Else if $\frac{1}{\sin\theta_1}<\gamma_1<\gamma_2<\frac{1}{\sin\theta_2}$, then
\begin{eqnarray}
\label{case3}
A(\delta) =
\begin{cases}
  \displaystyle g_{+}(\delta,\theta_1),  &  \delta_{min} \leq \delta < f_{\delta}(\gamma_1,\theta_1) \\
                \gamma_1, & f_{\delta}(\gamma_1,\theta_1) \leq \delta < f_{\delta}(\gamma_1,\theta_2) \\
                g_{-}(\delta,\theta_2), &  f_{\delta}(\gamma_1,\theta_2) \leq \delta \leq f_{\delta}(\gamma_2,\theta_2)
\end{cases}
\end{eqnarray}

  \item Else if $\gamma_1<\frac{1}{\sin\theta_1}<\frac{1}{\sin\theta_2}<\gamma_2$, then
\begin{eqnarray}
\label{case4}
A(\delta) =
\begin{cases}
  \displaystyle g_{+}(\delta,\theta_1),  &  \delta_{min} \leq \delta < f_{\delta}(\gamma_1,\theta_1) \\
                max[\gamma_1,g_{-}(\delta,\theta_2)], & f_{\delta}(\gamma_1,\theta_1) \leq \delta < \frac{1}{\sin\theta_1}\\
                max[\gamma_1,g_{-}(\delta,\theta_2)], &  \frac{1}{\sin\theta_1} \leq \delta \leq \frac{1}{\sin\theta_2}
\end{cases}
\end{eqnarray}

  \item Else if $\gamma_1<\frac{1}{\sin\theta_1}<\gamma_2<\frac{1}{\sin\theta_2}$, then
\begin{eqnarray}
\label{case5}
A(\delta) =
\begin{cases}
  \displaystyle g_{+}(\delta,\theta_1),  &  \delta_{min} \leq \delta < f_{\delta}(\gamma_1,\theta_1) \\
                \gamma_1, & f_{\delta}(\gamma_1,\theta_1) \leq \delta < \frac{1}{\sin\theta_1}\\
                max[\gamma_1,g_{-}(\delta,\theta_2)], &  \frac{1}{\sin\theta_1} \leq \delta \leq f_{\delta}(\gamma_2,\theta_2)
\end{cases}
\end{eqnarray}

\end{enumerate}

For Equation (\ref{case4}) and (\ref{case5}), what needs to be specifically noted is the situation when $f_{\delta}(\gamma_1,\theta_1) \leq \delta < \frac{1}{\sin\theta_1}$, the integral calculating $P_{\delta}(\delta)$ is the sum of two parts, i.e., $\int_{A(\delta)}^{A_1(\delta)}+\int_{A_2(\delta)}^{B(\delta)}$, and $A_1(\delta)=g_{-}(\delta,\theta_1)$, and $A_2(\delta)=g_{+}(\delta,\theta_1)$. Figure \ref{fA1} shows the PDF of $\delta$ with $P_{\gamma}(\gamma)\propto \gamma^{-1.5}$ for the five cases discussed above.

\section{The conditional probability distribution of $\log\mathcal{L}_{c}$ given $\log L_{c}$}\label{Acpd}
From Equation (\ref{delta_p}), we have
\begin{eqnarray}
\label{logdelta_p}
\log\mathcal{L}_{c}=\log L_{c}+q\log \delta.
\end{eqnarray}
If $q$ is a constant, according to the univariate theory of probability transformation, the conditional probability distribution of $\log\mathcal{L}_{c}$ given $\log L_{c}$ is
\begin{eqnarray}
\label{beamed_condition1}
p(\log \mathcal{L}_{c} \mid \log L_{c})= \frac{\ln10}{q}(\frac{\mathcal{L}_c}{L_c})^{\frac{1}{q}}P_{\delta}\left((\frac{\mathcal{L}_c}{L_c})^{\frac{1}{q}}\right).
\end{eqnarray}
Else if $q=q_c+\alpha$, where $q_c$ is a constant and $\alpha$ is the spectral index of radio core, $q$ will follow the similar distribution with $\alpha$. As mentioned in section \ref{tSID}, the distribution of $\alpha$ is well fitted by a Gaussian function with mean and sigma given in Table \ref{tab:sid}. Thus the PDF for $q$ is
\begin{eqnarray}
\label{P_p}
P_q(q)=\frac{1}{\sqrt{2\pi}\sigma}\exp\left(-\frac{(q-\mu)^2}{2\sigma^2}\right),
\end{eqnarray}
where $\sigma=0.397$ and $\mu=q_c+0.001$. Since $\log\mathcal{L}_{c}$ is the function of $q$ and $\delta$, the PDF for $\log\mathcal{L}_{c}$ is
\begin{eqnarray}
\label{P_p}
p(\log \mathcal{L}_{c})=\int P_{\delta}(\delta)P_q(q)\left|\frac{dq}{d\log\mathcal{L}_{c}}\right|d\delta.
\end{eqnarray}
From Equation (\ref{logdelta_p}), we have
\begin{eqnarray}
\label{P_p}
q=\frac{\log\mathcal{L}_{c}-\log L_{c}}{\log \delta},
\end{eqnarray}
Thus the conditional probability distribution of $\log\mathcal{L}_{c}$ given $\log L_{c}$ is
\begin{eqnarray}
\label{beamed_condition2}
p(\log \mathcal{L}_{c} \mid \log L_{c})=\int_{\delta_{min}}^{\delta_{max}}P_{\delta}(\delta)P_q\left(\frac{\log\mathcal{L}_{c}-\log L_{c}}{\log \delta}\right)\left| \frac{1}{\log \delta}\right|d\delta.
\end{eqnarray}

\listofchanges

\end{document}